# Transport response of topological hinge modes in $\alpha$-Bi$_4$Br$_4$

**Authors:** Md Shafayat Hossain[1]*†, Qi Zhang[1]*†, Zhiwei Wang[2,3]*, Nikhil Dhale[4]*, Wenhao Liu[4], Maksim Litskevich[1], Brian Casas[5], Nana Shumiya[1], Jia-Xin Yin[1], Tyler A. Cochran[1], Yongkai Li[2,3], Yu-Xiao Jiang[1], Ying Yang[2,3], Guangming Cheng[6], Zi-Jia Cheng[1], Xian P. Yang[1], Nan Yao[6], Titus Neupert[7], Luis Balicas[5], Yugui Yao[2,3]†, Bing Lv[4]†, M. Zahid Hasan[1,]†

**Affiliations:**

[1]Laboratory for Topological Quantum Matter and Advanced Spectroscopy (B7), Department of Physics, Princeton University, Princeton, New Jersey, USA.

[2]Centre for Quantum Physics, Key Laboratory of Advanced Optoelectronic Quantum Architecture and Measurement (MOE), School of Physics, Beijing Institute of Technology, Beijing 100081, China.

[3]Beijing National Laboratory for Condensed Matter Physics and Institute of Physics, Chinese Academy of Sciences, Beijing 100190, China.

[4]Department of Physics, University of Texas at Dallas, Richardson, Texas 75080, USA.

[5]National High Magnetic Field Laboratory, Tallahassee, Florida 32310, USA.

[6]Princeton Institute for Science and Technology of Materials, Princeton University, Princeton, NJ, USA.

[7]Department of Physics, University of Zurich, Winterthurerstrasse, Zurich, Switzerland.

†Corresponding authors, E-mail: mdsh@princeton.edu; qz9@princeton.edu; ygyao@bit.edu.cn; blv@utdallas.edu; mzhasan@princeton.edu.

*These authors contributed equally to this work.

**Abstract:**
**Electronic topological phases are typified by the conducting surface states that exist on the boundary of an insulating three-dimensional bulk. While the transport response of the two-dimensional surface states has been extensively studied, the response of the one-dimensional hinge modes has not been demonstrated. Here, we provide evidence for quantum transport in gapless topological hinge states existing within the insulating bulk and surface energy gaps in a layered topological insulator $\alpha$-Bi$_4$Br$_4$. Our magnetoresistance measurements reveal pronounced Aharonov–Bohm oscillations that are periodic in h/e (where h denotes Planck's constant and e is the electron charge). The observed periodicity evinces quantum interference of electrons circumnavigating around the hinges. We also demonstrate that the h/e oscillations evolve as a function of magnetic field orientation, following the interference paths along the hinge modes that are allowed by topology and symmetry. Our findings reveal the quantum transport response of topological hinge modes with both topological nature and quantum coherence, which can be eventually applied to the development of efficient quantum electronic devices.**

**Main text:**
Surface states in topological insulators[1-3] have attracted significant attention thanks to their potential to facilitate backscattering-free transport response, which holds great promise for the development of ultra-low dissipation quantum electronic devices. Despite the discovery of numerous topological materials since the early 2000s, the achievement of quantum coherent transport on a large scale, which is critical for leveraging topological properties, remains a significant technological challenge. While there have been reports of quantum coherence in engineered nanostructures[4-10], their strong topological phases reside in a bulk energy gap that is only locally present over a limited momentum region in the Brillouin zone[11]. As transport captures the electronic properties over the entire Brillouin zone, the transport characteristics in devices made from these strong topological insulators are dominated by electrons from bulk bands rather than electrons from the topological surface state[11]. Therefore, constructing extremely thin nanoribbons or very small nanostructures is necessary to achieve quantum coherence in strong topological insulators[4-10], limiting their potential for applications. To overcome these limitations, a new class of materials called higher-order topological insulators[12-15] offer a unique solution. In higher-order topological insulators, the topological electron channel propagates along the hinges of the sample, providing a strongly confined quantum channel due to the topology itself[12-16]. However, as with strong topological insulators, previous transport studies on higher-order topological insulators focused on systems in which non-topological conducting states are present as well[17], limiting the potential for quantum coherent transport. Here, we investigate a higher-order topological insulator, namely $\alpha$-$Bi_4Br_4$[16,18,19], which stands out for possessing a fully gapped energy spectrum in both the bulk and surface states throughout the entire Brillouin zone, providing a promising platform for a topological hinge transport channel that is free of non-topological conducting states. (Note that the $\alpha$ phase of $Bi_4Br_4$ is classified as a higher-order topological insulator[16,18,19], while the $\beta$ phase has been demonstrated to be a weak topological insulator[20-22].) And in $\alpha$-$Bi_4Br_4$, for the first time, we demonstrate the quantum coherent transport through the topological hinge modes.

A hallmark of quantum coherent transport is the manifestation of the Aharonov-Bohm interference[23]. The Aharonov-Bohm effect[23], predicted almost 60 years ago, describes a phenomenon where a quantum wave is split into two waves that go around a closed path and interfere under the influence of an electromagnetic potential. The resulting interference pattern is determined by the magnetic flux enclosed by the waves. In the case of electrons, such a quantum interference occurs if the conduction electrons remain phase coherent after completing closed trajectories, resulting in a periodic oscillation in electrical resistance with a characteristic period of the magnetic field $\Delta B = \Phi_0/S$, where $\Phi_0 = h/e$ is the flux quantum, $S$ is the area over which the electron trajectories remain phase coherent, $h$ is Planck's constant and $e$ is the electron charge. It is worth noting that, for bulk carriers, various sample-specific, impurity-dependent paths exist, whose interference results in a well-known phenomenon called universal conductance fluctuations[24], which exhibit aperiodic field dependence and are commonly observed in small metallic and semiconducting structures. In contrast, for topological conduction channels, all phase-coherent trajectories participating in the quantum interference enclose the same area perpendicular to the $B$ field, which is different from universal conductance fluctuations. Here, we present magnetoresistance traces from our $\alpha$-$Bi_4Br_4$ samples that unambiguously show $B$-periodic oscillations, the hallmark of the Aharonov-Bohm effect stemming from phase coherent carriers.

$\alpha$-$Bi_4Br_4$ possesses a quasi-one-dimensional crystal structure, exhibiting interlayer stacking along the $c$-axis with quasi-one-dimensional chains running along the $b$-axis within each layer[16,18,19,25,26]; see Section II of Methods and Extended Fig. 1 for details on the crystal structure. To investigate the quantum coherent transport in $\alpha$-$Bi_4Br_4$, we fabricated four-point probe devices using atomically thin flakes obtained through mechanical exfoliation from the crystal's (001) surface, which offers the largest surface area and aligns with the preferred plane of $\alpha$-$Bi_4Br_4$ (see Extended Fig. 1**c**). These flakes typically exhibit a rectangular shape, with a markedly greater length along a specific direction, which aligns with the $b$-axis as confirmed by our electron diffraction measurements on a relatively thicker flake (details in Section IV of Methods and Extended Fig. 2). In Fig. 1**a**, we present an optical microscopy image of a four-point probe device comprising four atomic layers. The length of the flake (displayed in Fig. 1**a**) along the *b*-axis, denoted as *w*, is approximately 2 $\mu m$. The thickness (denoted as *t*) of the flake with the same optical contrast was determined using atomic force microscopy[27] and is depicted in Fig. 1**b**. For detailed information on sample fabrication and characterization, we refer to Sections III and IV of the Methods, and Extended Figs. 2 and 3. To measure the four-terminal resistance of the sample, we employed a standard lock-in technique with a current of 10 nA.

Before discussing our transport results, it is crucial to understand the current trajectories in the four-atomic layer $\alpha$-Bi$_4$Br$_4$. In $\alpha$-Bi$_4$Br$_4$, the hinge-state profile depends on surface termination[16,19]. When uniformly cut along the (001), (00-1), (100), and (-100) surfaces, one hinge mode always resides at the top surface and another at the bottom[16]. However, in samples with an even number of layers, as in our four- and six-layer flakes, the inversion symmetry is broken, causing the pair of hinge states to reside on the same side surface, either the (100) or (-100)[16,19] planes This leads to the pair of hinge states propagating along one side of the sample, as schematically depicted in Fig. 1**c** and supported by scanning tunneling microscopy results (see Section VII in Methods, Extended Figs. 4 and 5, and ref.[19]). In the ideal three-dimensional $\alpha$-Bi$_4$Br$_4$, which is infinitely long along the *b*-direction, gapless, topological surface modes are present and protected by the $C_2$ rotation symmetry around the [010] axis, on the (010) and (0-10) surfaces. In an infinitely long sample, this mode allows the hinge states to connect in a way that preserves the $C_2$ symmetry since the hinge modes need to form a singly connected loop due to the spectral flow they carry. However, in finite-sized samples like our flakes, the (010) surface, being a crystal termination perpendicular to the atomic chains along the *b*-axis, inevitably features the absence of the $C_2$ symmetry. Consequently, the surface state at the (010) surface gaps out, but its "residue" forms the gapless hinge states, connecting the original hinge states to form a singly connected loop within the (100) surface carrying the spectral flow[16] (Fig. 1**c**). For more comprehensive discussions, we refer to Sections VIII and IX, and Extended Figs. 6 and 7 in the Methods.

Having discussed the formation of a hinge mode loop in the (100) surface, we focus on our transport results. Figure 1**d** captures the highlight of our experiments, illustrating the sample magnetoresistance when a magnetic field is applied perpendicularly to the *ab* plane at a very low temperature, $T \simeq 30$ mK. Examining the low-field resistance ($R_{xx}$) data, we detect a sharp cusp at zero field, followed by a concave-downward increase in $R_{xx}$ (*B*) at low magnetic fields $|B| < 1$ T). This behavior is characteristic of the weak anti-localization effect[28,29], which arises due to the strong spin-orbit interaction in $\alpha$-Bi$_4$Br$_4$. As we increase the magnetic field strength, we observe pronounced and reproducible resistance oscillations with a period of $\Delta B \simeq 1.97$ T throughout the entire field range of $\pm 18$ T (Fig. 1**d** left inset). These *B*-periodic oscillations are a clear manifestation of the $h/e$ Aharonov-Bohm effect[23], where the period corresponds to an area of $\simeq 2.1 \times 10^{-15}$ m$^2$. Remarkably, this result is in agreement with the flux area of the sample's *bc* plane, $\left(\frac{\Phi}{B} \perp ab\right)_{bc} = w.t\, cos(107°) \simeq 2.3 \times 10^{-15}$ m$^2$ (see Fig. 1**g**), where $107°$ is the angle between the crystallographic *a* and *c* directions according to its crystal structure[16,18,19,25,26]. Note that the slight mismatch between the extracted area from the Aharonov-Bohm effect and $\left(\frac{\Phi}{B} \perp ab\right)_{bc}$ can likely be attributed to the experimental error bars in determining both *w* (via optical microscopy) and *t* (via atomic force microscopy), as well as the exact localization profile of the hinge state, which may not be precisely located at atomic positions right at the edge.

Further analysis of the Aharonov-Bohm effect provides crucial insights into our experiments. Our observations reveal that the amplitude of the oscillations is of the order of $\sim 1$ k$\Omega \simeq 0.04\, h/e^2$, which is comparable to those observed in quasi-one-dimensional nanostructures[4,8,30,31]. To obtain a more detailed view of the prominent *B*- periodic oscillation and its harmonics, we take the Fourier transform of the derivative d$R$/d$B$, as shown in the right inset of Fig. 1**d**. This method is commonly used to isolate the oscillatory part from the slowly varying background[31]. The resulting Fourier transform appears to be reasonably clean, indicating the absence of contributions from the bulk states and universal conductance fluctuations. If due to time-reversal symmetry, wave packets do not interfere after once, but only after twice encircling the hinge loop, $h/2e$ periodicity oscillations, known as the Altshuler–Aronov–Spivak oscillation[32,33], should be observed. Our Fourier transform result indeed reveals the presence of an $h/2e$ peak, subdominant compared to the $h/e$ Aharonov-Bohm peak, suggesting that the weak anti-localization effect, observed at $|B| < 1$ T, extends to sufficiently high field values. An overview of previous studies on the Aharonov-Bohm effect in mesoscopic transport experiments, encompassing results from metallic rings, cylinders, nanostructures of three-dimensional first and higher-order topological insulators, and a comparison of our observations with these previous results, is detailed in Section XI of the Methods.

In order to confirm that the observed Aharonov-Bohm oscillations stem from the hinge state carriers, we performed angular-dependent measurements. Figure 1**e** displays a series of magnetoresistance traces as the magnetic field direction is rotated from perpendicular to the *ab* plane to parallel to the *b* axis. The oscillation period D$B$ increases as the magnetic field is oriented towards the *b* axis. To better visualize this evolution, we plot the Fourier transform frequency (equivalent to $1/\Delta B$) as a function of the angle ($\theta$) between the magnetic field and the direction perpendicular to the *ab* plane (Fig. 1**f**). The main peak or frequency in the Fourier transform exhibits a clear $cos\theta$ angular dependence, which aligns with the phase-coherent hinge mode propagation scenario depicted in Fig. 1**c**.

To bolster the case for the Aharonov-Bohm effect arising from phase-coherent carriers circumnavigating along the hinges on the *bc* plane, we examined the magnetoresistance with the magnetic field oriented along different directions relative to the *a*-axis of the crystal. As illustrated in Fig. 2**a** for the hinge state propagation configuration expected for an even-layer $\alpha$-Bi$_4$Br$_4$, the magnetic flux should be at its maximum when $\delta = 17^o$ and follow a $\cos(\delta - 17^o)$ behavior as the field direction rotates away from *a*-axis towards an orientation $\perp ab$, with $\delta$ being the angle between the magnetic field and the *a*-axis. In Figs. 2**c-f**, we present the angular-dependent data obtained from a second four-layer $\alpha$-Bi$_4$Br$_4$ sample, Sample 2 (see Fig. 2**b** for its optical microscopy image). This sample has a length along the *b*-axis of approximately $w \simeq 2.1$ µm and a thickness $t = 3.87$ nm, as determined from a flake with the same optical contrast. We observe clear Aharonov-Bohm oscillations for both $B \parallel a$ (Fig. 2**c**) and $B \perp ab$ (Fig. 2**d**) orientations. For $B \parallel a$, we find oscillations with a periodicity of $\Delta B \simeq 0.52$ T, corresponding to an area of $7.9 \times 10^{-15}$ m$^2$ (Fig. 2**c**), while for $B \perp ab$, the periodicity is $\Delta B \simeq 1.92$ T, corresponding to an area of $2.2 \times 10^{-15}$ m$^2$ (Fig. 2**d**). These results are consistent with the respective flux areas or projections of the sample's *bc* plane, $\left(\frac{\Phi}{B} \parallel a\right)_{bc} = w.t\, sin(107^o) \simeq 7.8 \times 10^{-15}$ m$^2$ and $\left(\frac{\Phi}{B} \perp ab\right)_{bc} = w.t\, cos(107^o) \simeq 2.4 \times 10^{-15}$ m$^2$. Therefore, in this case, the Aharonov-Bohm effect for $B \parallel a$ exhibits a periodicity that is approximately 3.6 times smaller, indicating a flux area that is approximately 3.6 times larger when compared to that for $B \perp ab$. This value aligns relatively well with $\frac{\left(\frac{\Phi}{B} \parallel a\right)_{bc}}{\left(\frac{\Phi}{B} \perp ab\right)_{bc}} = \frac{sin(107^o)}{cos(107^o)} \simeq 3.3$, as expected for the hinge state propagation scenario shown in Fig. 2**a**. A more compelling indication of such hinge mode propagation is revealed in Figs. 2**e-f**, where we explore the Aharonov-Bohm effect at different angles ($\delta$) between the magnetic field and the *a*-axis. Figure 2**e** displays a series of magnetoresistance traces acquired at different $\delta$, ranging from $0^o$ ($B \parallel a$) to $90^o$ ($B \perp ab$). The period of the oscillations in field D$B$ decreases until $\delta \sim 17^o$ is reached, before increasing again as the field direction approaches $\perp ab$. To provide a clearer visualization of this evolution, we plot the Fourier transform frequency (equivalent to $1/\Delta B$) as a function of $\delta$ (Fig. 2**f**). Fitting the data in Fig. 2**f** to a cosine function yields a fitting function of $1.99 \cos(\delta - (16.4 \pm 0.3)^o)$, indicating that $1/\Delta B$, i.e., the flux area extracted from the Aharonov-Bohm effect, reaches a maximum around $(16.4 \pm 0.3)^o$. This observation closely aligns with the depicted phase-coherent hinge mode propagation scenario in Fig. 2**a**. Collectively, the angular dependencies presented in Figs. 1**f** and 2**f** provide compelling evidence indicating that the Aharonov-Bohm effect indeed arises from phase-coherent carriers circumnavigating along the hinges located on the *bc* plane of the sample.

After establishing the origin of the Aharonov-Bohm effect in $\alpha$-Bi$_4$Br$_4$, we explore its temperature dependence. Figure 3**a** displays the temperature evolution of the magnetoresistance oscillations in Sample 2. The oscillatory features persist up to $T \simeq 20$ K, beyond which the resistance decreases considerably (as also observed in our temperature-dependent resistance measurements detailed in Section XII and Extended Fig. 8 in the Methods), presumably due to the thermal excitation of the bulk carriers towards the conduction band. To summarize the temperature-dependent data, we take the Fourier transform of d$R$/d$B$ and plot the Fourier transform magnitude of the primary $h/e$ peak against temperature in Fig. 3**b**. We find that the amplitude of the Aharonov-Bohm oscillations follows a clear power law as function of the temperature, i.e., $T^\alpha$, with the exponent a being very close to the canonical value $\alpha = -1/2$ observed in nearly all previous experiments probing the Aharonov-Bohm effect[34-36] including mesoscopic metal rings[34] and strong topological insulator nanoribbons[4]. Note that the amplitude of the Altshuler–Aronov–Spivak oscillation also follows a power law as function of the temperature (Fig. 3**c**) with $\alpha$ reasonably close to $-1/2$. However, in certain experiments the amplitude of the Aharonov-Bohm oscillations is found to instead follow a nearly exponential decay as a function $T$ (ref. [10]). This was interpreted as a temperature dependent phase coherence length, implying a modified expression for the amplitude of the Aharonov-Bohm oscillations: $\Delta G = \frac{e^2}{h} \left( \frac{2\pi h D}{L_\phi^2 k_B T} \right)^{1/2} \exp(-\pi P/L_\phi)$ [ref. [35]]. Here, $D$ is the diffusion constant, $e$, $h$, and $k_B$ are the electron charge, Planck, and Boltzmann constants, and $L_\phi$ is the phase coherence length, and $P$ represents the perimeter of the geometrical path enclosing the magnetic flux. In this expression, at least two functional forms were proposed for $L_\phi$, i.e., either $L_\phi \propto T^{-\frac{1}{2}}$ (ref. [35]), or $L_\phi \propto T^{-1}$ (ref. [10]). In Extended Fig. 9, we provide examples of fits of our Aharonov-Bohm and Altshuler–Aronov–Spivak oscillations to the above expression. We point out that this expression leads to a poorer description of our experimental data, relative to the simple power law, regardless of the precise functional form for $L_\phi$. This suggests that in our devices, $L_\phi$ is probably $T$-independent. As the fitting indicates, regardless of its exact $T$-dependence, $L_\phi$ is always much longer than the perimeter enclosed by the hinge modes in our four-layer a-Bi$_4$Br$_4$ device. Additionally, based on our data in Figs. 1 and 2, showing a close correspondence between the area over which the electron trajectories remain phase coherent, as determined from the Aharonov-Bohm oscillations, and the respective flux areas of the sample's *bc* plane, we can infer the phase-coherent diffusion length for our $\alpha$-Bi$_4$Br$_4$ sample. It is at least as large as the perimeter of the sample's *bc* plane, i.e., $2(w + t) \simeq 4.2$ μm.

In Figs. 1-3, we have presented data for two devices with similar physical dimensions, yielding comparable periods for the Aharonov-Bohm oscillations for specific magnetic field orientations. However, to firmly establish the correlation between the oscillation period and the area enclosed by the phase coherent carriers circumnavigating the hinges in the *bc* plane, it is beneficial to examine another device with different dimensions along the *bc* plane. Thus, in Fig. 4, we present magnetoresistance data obtained from a six-layer $\alpha$-Bi$_4$Br$_4$ sample, denoted as Sample 3. Figure 4**a** portrays its optical microscopy image. This sample has a length along the *b*-axis of approximately $w \simeq 2.2$ μm and a thickness $t = 5.75$ nm, as determined from a flake with the same optical contrast (Fig. 4**b**). Notably, the hinge state propagation configuration in the six-layer flake, as schematically illustrated in Fig. 4**c**, mirrors that of the four-layer (and any even-layer) flake. Akin to Sample 2 (Fig. 2), we observe Aharonov-Bohm oscillations having distinct periodicities for both $B \perp ab$ (Fig. 4**d**) and $B \parallel a$ (Fig. 4**e**) orientations. For $B \perp ab$, we find oscillations with a periodicity $\Delta B \simeq 1.25$ T, corresponding to an area of $3.3 \times 10^{-15}$ m² (Fig. 4**d**), while for $B \parallel a$, the periodicity is $\Delta B \simeq 0.39$ T, corresponding to an area of $10.6 \times 10^{-15}$ m² (Fig. 4**e**). These results reasonably match the respective flux areas of the sample's $bc$ plane, $\left(\frac{\Phi}{B} \perp ab\right)_{bc} = w.t\cos(107º) \simeq 3.7 \times 10^{-15}$ m² and $\left(\frac{\Phi}{B} \parallel a\right)_{bc} = w.t\sin(107º) \simeq 12 \times 10^{-15}$ m². Therefore, here also, the Aharonov-Bohm effect for $B \parallel a$ exhibits a periodicity that is approximately 3.2 times smaller, indicating a flux area approximately 3.2 times larger when compared to that for the $B \perp ab$ case. This observation closely aligns with $\frac{\left(\frac{\Phi}{B}\parallel a\right)_{bc}}{\left(\frac{\Phi}{B}\perp ab\right)_{bc}} = \frac{sin(107º)}{cos(107º)} \simeq 3.3$, as anticipated according to the hinge state propagation scenario displayed in Fig. 4**c**. Thus, the close correspondence between the period of the Aharonov-Bohm oscillations and the respective flux areas of the sample's $bc$ plane observed in samples with different physical dimensions robustly establishes the relationship between the period of the oscillations and the area enclosed by phase coherent carriers circumnavigating along the hinges on the $bc$ plane.

Before closing, it is worth mentioning that our investigation has primarily focused on even-layer samples hosting hinge states that propagate along one side of the sample. The odd layer samples, on the other hand, are theoretically envisioned to exhibit inversion-symmetric helical hinge modes with a different propagation configuration[16]. Their intriguing transport response awaits future investigations.

In conclusion, our experiments provide the first compelling evidence for the existence of quantum coherence in topological hinge modes, thus opening new avenues towards the development of integrated topological circuitry. Unlike conventional electronic devices, topological circuits are robust against defects and impurities, making them far less prone to energy dissipation which is advantageous for practical applications. Our work goes beyond spectroscopic probes and showcases the potential of transport experiments to probe topological quantum matter, representing an important step towards realizing functional devices based on higher-order topology. Importantly, the nearly non-dissipative transport, implied by our observation of the quantum interference effect, hints at the potential of topological insulators with large topological gaps to serve as superior materials for interconnects among semiconducting elements. Overall, our work demonstrates the rich physics and technological potential of topological insulators and could pave the way towards novel applications in quantum information processing and spintronics.

**Acknowledgement:** We acknowledge C. Yoon and F. Zhang for providing the calculated hinge states. Experimental and theoretical work at Princeton University was supported by the Gordon and 286 Betty Moore Foundation (GBMF4547; M.Z.H.). The material characterization is supported by the United States 287 Department of Energy (US DOE) under the Basic Energy Sciences program (grant number DOE/BES DE-FG-288 02-05ER46200; M.Z.H.). L.B. is supported by DOE-BES through award DE-SC0002613. The National High Magnetic Field Laboratory (NHMFL) acknowledges support from the US-NSF Cooperative agreement Grant number DMR-1644779 and the state of Florida. We thank T. Murphy, G. Jones, L. Jiao, and R. Nowell at NHMFL for technical support. T.N. acknowledges supports from the European Union's Horizon 2020 research and innovation programme (ERC-StG-Neupert-757867-PARATOP). Crystal growth at University of Texas at Dallas is supported by US Air Force Office of Scientific Research Grant No. FA9550-19-1-0037 and National Science Foundation (NSF)-DMREF- 1921581; B.L. Crystal growth at Beijing Institute of Technology is funded by the National Science

Foundation of China (NSFC) (11734003; Z.W.), the National Key Research and Development Program of China (2016YFA0300600; Z.W.). Y.G.Y. is supported by NSFC (11574029) and the Strategic Priority Research Program of Chinese Academy of Sciences (XDB30000000).

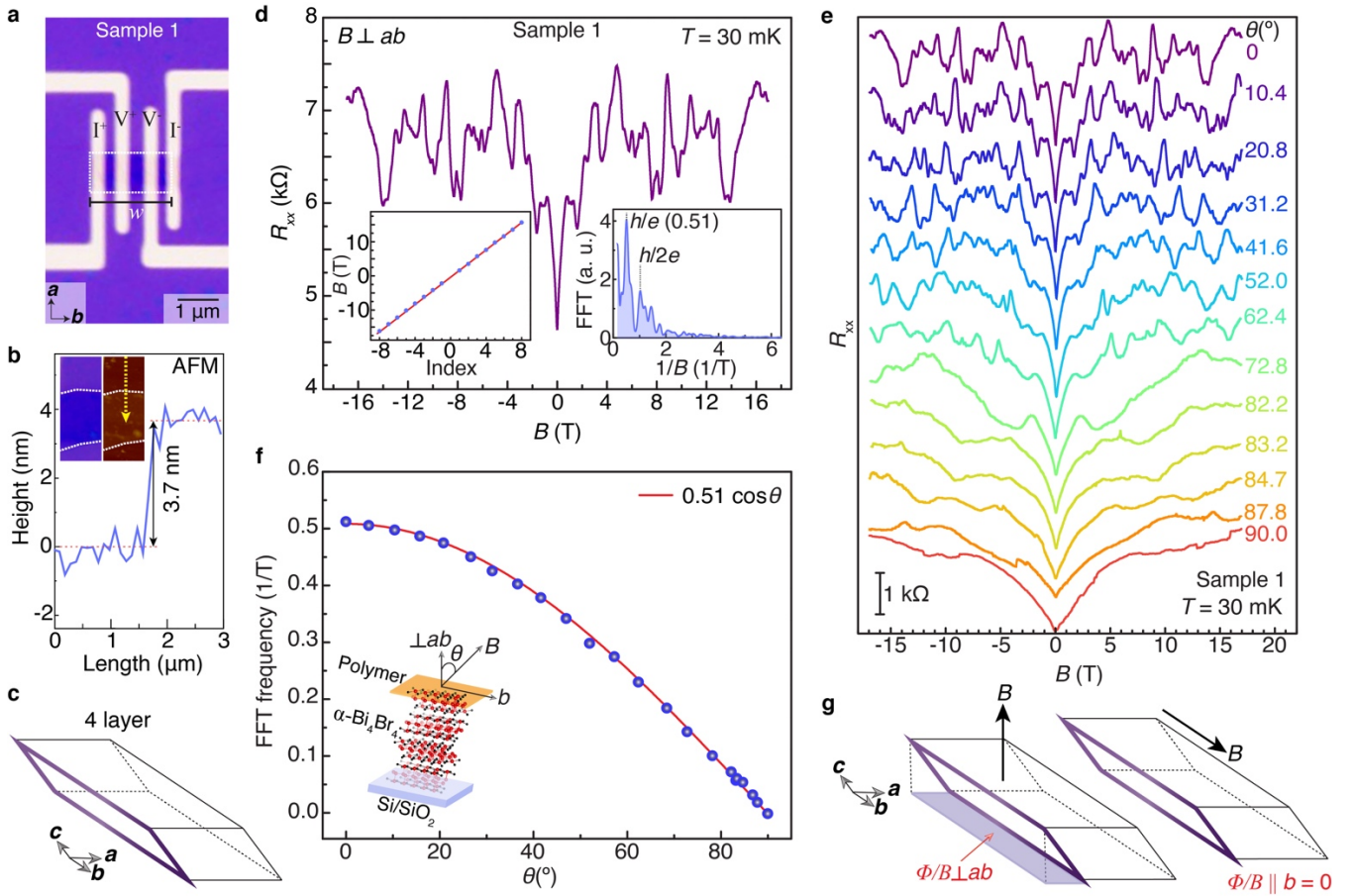

**Fig. 1: Observation of Aharonov-Bohm oscillations of the topological hinge states. a,** Optical microscopy image of Sample 1: A mechanically exfoliated $\alpha$-Bi$_4$Br$_4$ flake (before encapsulating it with a polymethyl methacrylate thin film) comprising four atomic layers. White dashed lines indicate the flake boundaries. Current ($I^+$ and $I^-$) and voltage ($V^+$ and $V^-$) probes for four-terminal electrical transport measurements are also shown. The length of the flake along the *b*-axis, denoted as *w*, is also marked. **b,** Atomic force microscopy image of a $\alpha$-Bi$_4$Br$_4$ flake (right inset) having the same optical contrast as the flake in panel **a**, with its height (*H*) profile collected along the yellow dashed line (the arrow marks the directions of the scan), indicating a thickness $t = H/sin(107°) = 3.87$ nm which matches the expected thickness for four layers. The flake boundaries are marked with white dashed lines. Left inset: Optical microscopy image. **c,** Theoretically envisioned propagation of the topological hinge states in a four-layer Bi$_4$Br$_4$ sample, where the breaking of the inversion symmetry leads to hinge state propagation along one side of the sample.

The inversion asymmetric path for the helical hinge states when the $C_2$ symmetry is broken on the *ac* plane is highlighted in purple. **d**, Magnetoresistance of the Bi$_4$Br$_4$ device with the magnetic field applied perpendicularly to the *ab* plane. The magnetoresistance trace features a clear modulation. Left inset: Magnetic field positions of the prominent resistance minima (highlighted in circles) that can be well fitted to a straight line of slope (1.97±0.005) T, signaling a *B*-periodic oscillation. Right inset: Fourier transform of the derivative d*R*/d*B* over the entire field range, showing a well-developed peak at 0.51 1/T, denoting a periodicity of (1/0.51) = 1.96 T. Locations of the peaks attributed to *h*/*e* and *h*/2*e* magnetic flux threaded into the area perpendicular to the magnetic field are labeled. **e**, Angular dependence of the Aharonov-Bohm oscillations as the direction of the magnetic field is varied from perpendicularly to the *ab* plane ($\theta = 0°$) to along the *b*-axis ($\theta = 90°$). Plotted are a set of magnetoresistance traces collected at different angles ($\theta$) between the magnetic field and the direction perpendicular to the *ab* plane. **f**, Fourier transform frequency (in units of 1/T) as a function of the angle between the magnetic field and the direction perpendicular to the *ab* plane. The angular dependence follows a cos$\theta$ behavior, indicating the two-dimensionality of the area encompassed by the quantum coherent current propagation. The inset illustrates the direction of rotation. **g**, Schematic depiction of the projected areas (shown as the shaded purple region), where the magnetic flux is threaded into for magnetic fields perpendicular to the *ab* plane (left) and along the *b* axis (right), and which are enclosed by the topological hinge state propagation (shown in panel **c**) in a four-layer Bi$_4$Br$_4$ sample. The conduction path for the helical hinge states is highlighted in purple. Within this hinge state propagation scenario, during the rotation of the magnetic field direction from ⊥ *ab* to || *b*, the magnetic flux becomes maximum for *B* ⊥ *ab* and decreases to zero for *B* || *b*, following a cos$\theta$ behavior that is consistent with the experimental data.

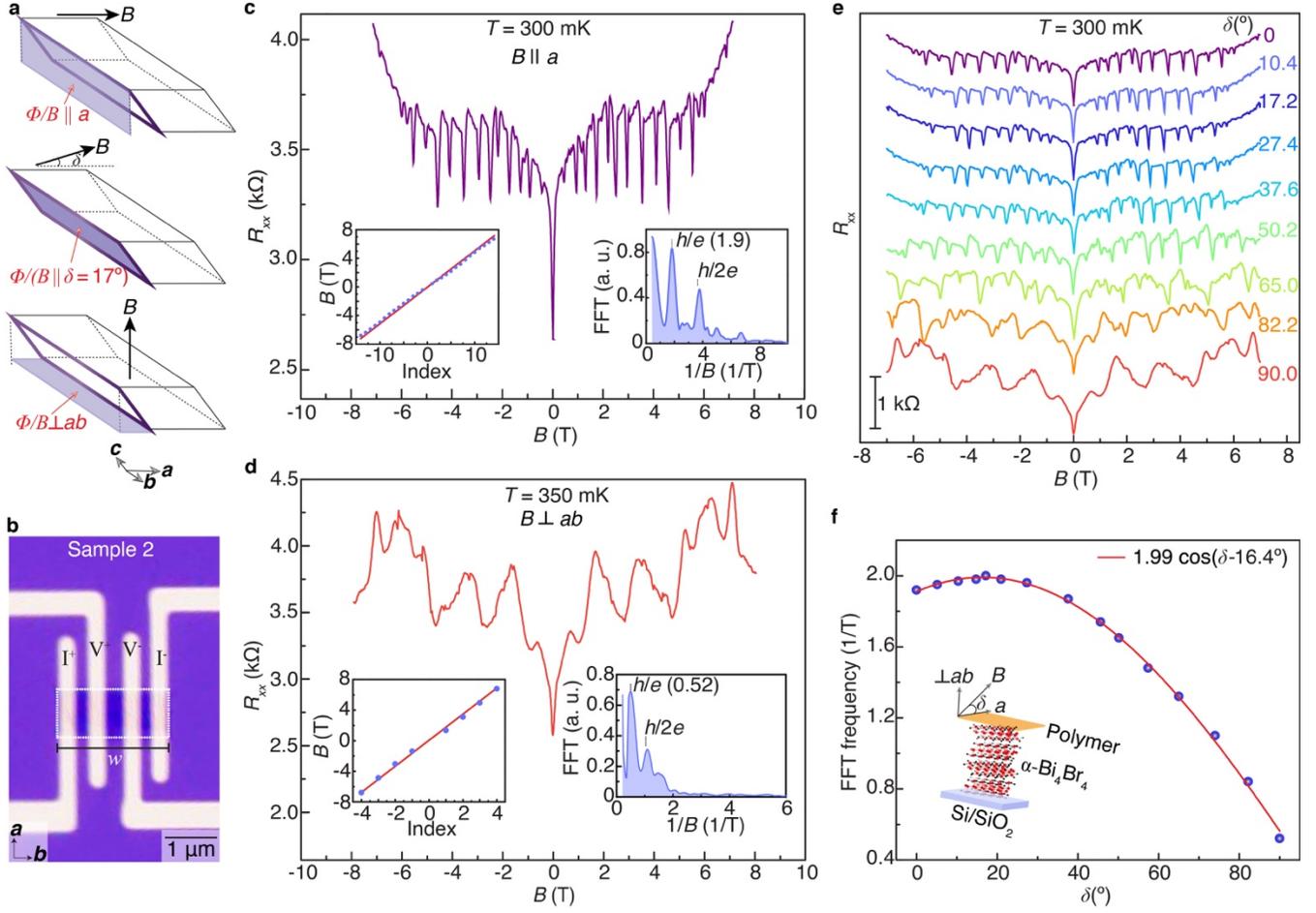

**Fig. 2: Angular dependence of the Aharonov-Bohm oscillations confirming hinge state formation on the *bc* plane. a,** Schematic illustration of the projected areas enclosed by propagating hinge states (Fig. 1**c**) in a four-layer α-Bi$_4$Br$_4$ sample (shown as shaded purple regions), through which the magnetic flux penetrates into when the magnetic fields are applied along the *a*-axis (top), rotated by 17° relative to the *a*-axis (middle), and perpendicularly to the *ab* plane (bottom), leading to Aharonov-Bohm oscillations . The conduction path for the helical hinge states is highlighted by purple line. In this hinge state propagation configuration, during the rotation of the magnetic field direction from $\parallel a$ to $\perp ab$, the magnetic flux is maximum for $\delta = 17°$ and follows a $\cos(\delta - 17°)$ behavior, where $\delta$ is the angle between the magnetic field and the *a*-axis. **b,** Optical microscopy image of Sample 2: Another mechanically exfoliated α-Bi$_4$Br$_4$ flake (prior to encapsulation with a polymethyl methacrylate thin film) also comprising four atomic layers. White dashed lines indicate the flake boundaries. Current ($I^+$ and $I^-$) and voltage ($V^+$ and $V^-$) probes for four-terminal electrical transport measurements are also shown. The length of the flake along the *b*-axis, denoted as *w*, is also indicated. **c,** Resistance as a function of the magnetic field for Sample 2, with the magnetic field applied parallel to the *a*-axis, revealing a pronounced, periodic modulation. Left inset: Values of magnetic field for which the prominent resistance minima are observed, that can be well fitted to a straight line yielding a slope of $(0.52 \pm 0.008)$ T, indicating *B*-periodic oscillations. Right inset: Fourier transform of d*R*/d*B* over the entire field range, exhibiting a well-developed peak at 1.9 1/T, corresponding to a periodicity of $(1/1.9) = 0.53$ T. Locations of the peaks corresponding to *h/e* and *h/2e* are labeled. **d,** Magnetoresistance measurements from the same sample but with the magnetic field perpendicular to the *ab* plane, revealing a distinct modulation pattern that is more sparsely spaced in *B* when compared to the data in panel **c**. Left inset: Values of magnetic field for which prominent resistance minima are observed that can be fitted to a straight line with slope (1.92±0.015) T, signaling a *B*-periodic oscillation. Right inset: Fourier transform of d*R*/d*B* over the entire field range, revealing a well-developed peak at 0.52 1/T, suggesting a periodicity of (1/0.52) = 1.92 T. Peaks corresponding to *h/e* and *h/2e* are labeled. Therefore, the Aharonov-Bohm effect for $B \parallel a$ yields a periodicity that is approximately 3.6 times smaller, indicating a flux area that is approximately 3.6 times larger when compared to that for $B \perp ab$. This observation reasonably matches the expected value of $\left(\frac{\Phi}{B\parallel a}\right)_{bc} / \left(\frac{\Phi}{B} \perp ab\right)_{bc} = \frac{sin(107°)}{cos(107°)} \simeq 3.3$. **e**, Angular dependence of the Aharonov-Bohm oscillations as a function of magnetic field orientation, i.e., from along the *a*-axis ($\delta = 0°$) to perpendicular to the *ab* plane ($\delta = 90°$). Plotted are a set of magnetoresistance traces collected at different angles ($\delta$) between the magnetic field and the *a*-axis. **f**, Fourier transform frequency (in units of 1/T) plotted as a function of the angle between the magnetic field and the *a*-axis. The angular dependence can be fitted well to $cos(\delta - (16.4 \pm 0.3)°)$, providing strong evidence for hinge states on the *bc* plane of the sample, as displayed in panel **a**.

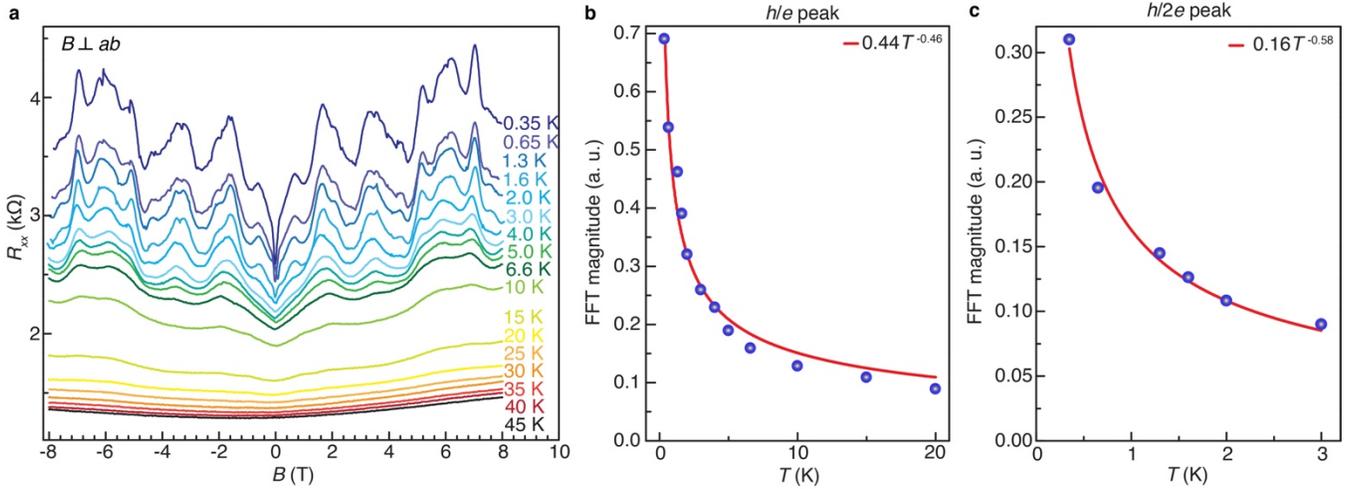

**Fig. 3: Temperature dependence of Aharonov-Bohm oscillations. a**, Magnetoresistance traces (for $B \perp ab$) at different temperatures, ranging from 0.35 K to 45 K, acquired from Sample 2, showing the temperature evolution of the amplitude of the Aharonov-Bohm oscillations. **b**, Temperature dependence of the Fourier transform amplitude of the primary *h/e* oscillations. Fitting the data to $AT^x$ yields the fitting parameters $A = 0.44$ and $x = -0.46$. The resulting fitting function is represented by the red curve. The fitting curve, which follows the form $0.44\, T^{-0.46\pm0.03}$, aligns well with the expected $T^{-0.5}$ relation for Aharonov-Bohm oscillations. **c**, Temperature dependence of the Fourier transform amplitude of the *h/2e* oscillations. Fitting the data to $AT^x$ yields the fitting parameters $A = 0.16$ and $x = -0.58$. The resulting fitting function is represented by the red curve. The fitting curve, which follows the form $0.16\, T^{-0.58\pm0.03}$, in good agreement with the canonical $T^{-0.5}$ relationship.

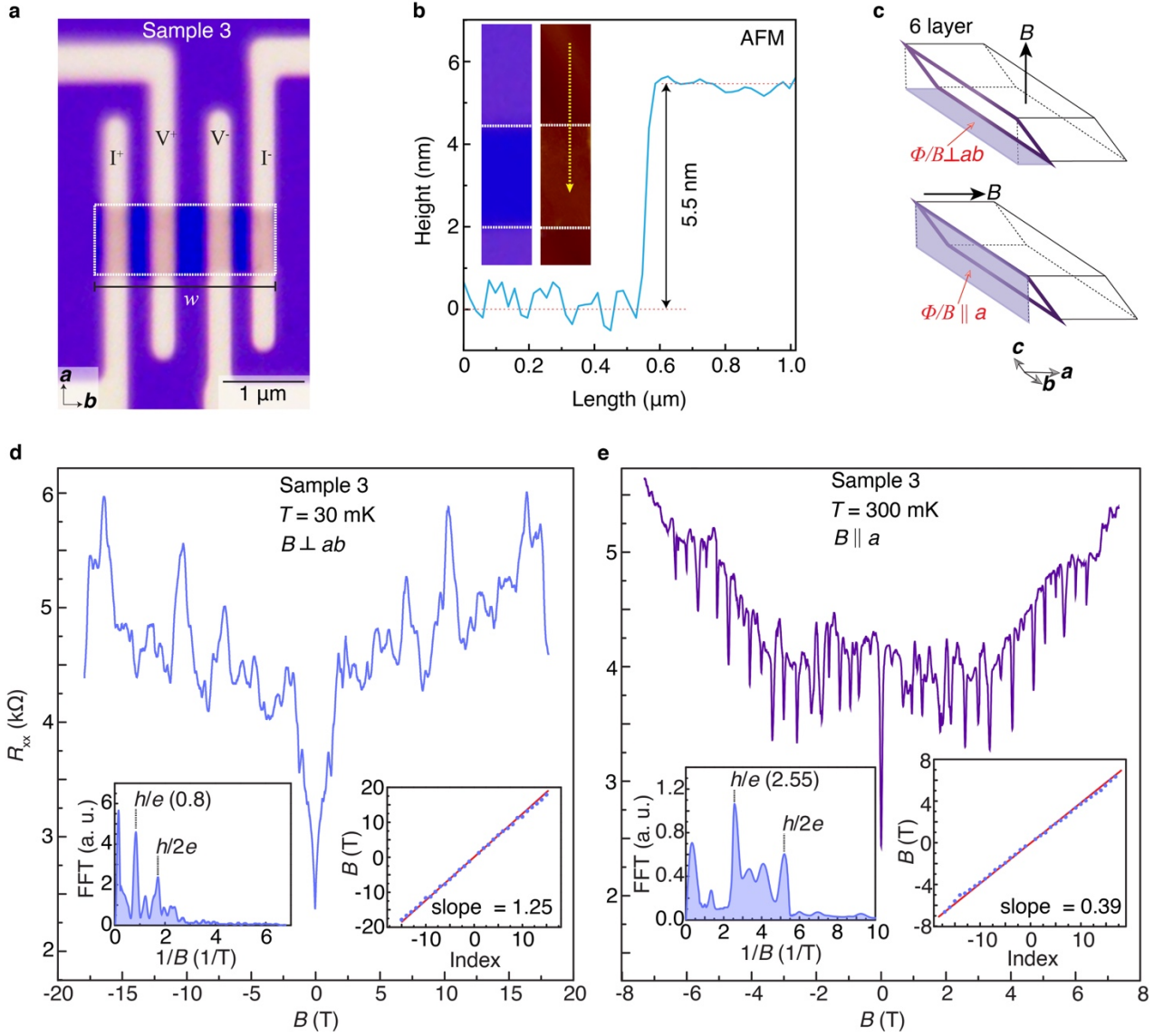

**Fig. 4: Observation of Aharonov-Bohm oscillations in a six-layer $\alpha$-Bi$_4$Br$_4$ sample. a,** Optical microscopy image of Sample 3: A mechanically exfoliated $\alpha$-Bi$_4$Br$_4$ flake (before encapsulating it with a polymethyl methacrylate thin film) comprising six atomic layers. White dashed lines indicate the flake boundaries. Current ($I^+$ and $I^-$) and voltage ($V^+$ and $V^-$) probes for four-terminal electrical transport measurements are also shown. The length of the flake along the *b*-axis, denoted as $w$, is also marked. **b,** Atomic force microscopy image of a $\alpha$-Bi$_4$Br$_4$ flake (right inset). Its optical contrast matches that of the flake in panel **a**. Its height ($H$) profile collected along the yellow dashed line (the arrow marks the directions of the scan), indicates a thickness $t = H/sin(107^o) = 5.75$ nm which reasonably matches the expected thickness for six layers. The flake boundaries are marked with white dashed lines. Left inset: corresponding optical microscopy image. **c,** Schematic depiction of the area (shown as the shaded purple region), where the magnetic flux is threaded into for magnetic fields applied perpendicularly to the *ab* plane (top) and along the *a*-axis (bottom), that are enclosed by the topological hinge state propagating (shown in Fig. 1**c**) in a six-layer $\alpha$-Bi$_4$Br$_4$ sample. The conduction path for the helical hinge states is highlighted in purple. This hinge state propagation configuration is akin to the one expected for the four-layer case, as depicted in Figs. 1 and 2. **d,** Resistance $R_{xx}$ as a function of the magnetic field in Sample 3, with the field applied perpendicularly to the *ab* plane.

$R_{xx}$ reveals pronounced, periodic modulations. Left inset: Positions of the prominent resistance minima as function of magnetic field. These can be well fitted to a straight line of slope $s = (1.25 \pm 0.01)$ T, indicating $B$-periodic oscillations. Right inset: Fourier transform of $dR/dB$ over the entire field range, exhibiting a well-developed peak at 0.8 1/T, corresponding to a periodicity of $(1/0.8) = 1.25$ T. The locations of the peaks corresponding to $h/e$ and $h/2e$ are labeled. **e**, Magnetoresistance measurements on the same sample but with the magnetic field parallel to the $a$-axis, unveiling a distinct modulation pattern that is more densely spaced in $B$ when compared to the data presented in panel **d**. Left inset: Magnetic field positions of the prominent resistance minima that can be well fitted to a straight line having a slope $s = (0.39 \pm 0.006)$ T, indicating $B$-periodic oscillations. Right inset: Fourier transform of $dR/dB$ over the entire field range, exhibiting a well-developed peak at 2.55 1/T, corresponding to a periodicity of $(1/2.55) = 0.39$ T. The locations of the peaks corresponding to $h/e$ and $h/2e$ are labeled. Therefore, akin to the four-layer case, here the Aharonov-Bohm effect for $B \parallel a$ yields a periodicity approximately 3.2 times smaller, indicating a flux area approximately 3.2 times larger when compared to the $B \perp ab$ case. This observation closely matches the expected value of $\left(\frac{\Phi}{B \parallel a}\right)_{bc} / \left(\frac{\Phi}{B} \perp ab\right)_{bc} = \frac{sin(107º)}{cos(107º)} \simeq 3.3$.

**Methods:**

I. **Single crystal synthesis**

The single crystals were synthesized using solid state reactions. The Bi (Alfa Aesar, 99.9999%) pieces together with $HgBr_2$ powders (Alfa Aesar, 99% +) were measured in a stoichiometric Bi: Br = 1:1 ratio, was sealed in an evacuated quartz tube. The tube was then put into a horizontal two zone tube furnace with raw materials was placed at the hot end at 265 °C and cold end was set at 210 °C. Black needle-shaped crystals start to form at the cold end and in the middle of the quartz tube after a few days. The whole assembly was then further annealed in a low temperature oven at 160 °C for over one month to ensure the high quality of the grown crystals after the full two weeks' reaction. Large crystals up to 6 × 0.7 × 0.5 mm³ with needle direction along $b$ axis and large lateral area along (*00l*) surface could be grown using this method. All the materials handling and processing were performed inside purified Ar-atmosphere glovebox with total $O_2$ and $H_2O$ levels < 0.1 ppm. The structure of the single crystals is confirmed by both X-ray single crystal diffraction using a Bruker Apex DUO single crystal diffractometer and Rigaku Smartlab X-ray diffractometer, and the composition is determined by SEM energy-dispersive X-ray spectroscopy (SEM-EDX) using Zeiss EVO LS 15 SEM with accelerating voltage of 20 keV.

II. **Crystal structure of *α*-Bi₄Br₄**

*α*-$Bi_4Br_4$ possesses a quasi-one-dimensional crystal structure[16,18,19,25,26]. The side view of the crystal, as shown in Extended Fig. 1**a**, reveals an AB interlayer stacking along the $c$ axis. In Extended Fig. 1**b**, the top view illustrates the quasi-one-dimensional chains running along the $b$ axis within each layer. The lattice in the $ab$ monolayer breaks the mirror symmetry along the $a$-axis, distinguishing the A-type layer from the B-type layer. The determination of this lattice's mirror symmetry breaking is pivotal for analyzing step edge geometry and understanding the topology in *α*-$Bi_4Br_4$, which we present in Extended Figs. 4 and 5. Notably, adjacent layers exhibit opposite tilts, confirmed by scanning transmission electron microscopy measurements[19]. To investigate the crystal structure of *α*-$Bi_4Br_4$, we conducted single-crystal X-ray diffraction using a Bruker SMAER Apex II X-ray diffractometer with a Mo Kα source (Extended Fig. 1**c**). The resulting lattice parameters are: $a = 13.0667(12)$ Å, $b = 4.3359(4)$ Å, $c = 20.0676(19)$ Å, and $\beta = 107.336(2)º$. The lattice parameters obtained from our X-ray diffraction results quantitatively match prior experimental findings (e.g., ref. [18,26]) and align well with the theoretically predicted values reported in ref. [16]. We also confirmed the preferred plane to be (001) using a Rigaku Smartlab diffractometer with Cu Kα radiation. It is worth noting that we conducted mechanical exfoliation specifically on this (001) surface to acquire the atomically thin flakes used for our transport experiments.

### III. Device fabrication

We employed a polydimethylsiloxane (PDMS) stamp-based mechanical exfoliation technique to fabricate atomically thin (four- or six-atomic-layer-thick) $\alpha$-Bi$_4$Br$_4$ devices. We patterned the sample contacts on the silicon substrates with a 280 nm layer of thermal oxide using electron beam lithography, followed by chemical development and metal deposition (5 nm Cr/35 nm Au). The fresh $\alpha$-Bi$_4$Br$_4$ flakes were mechanically exfoliated from bulk single crystals on PDMS stamps. Prior to transferring them onto the SiO$_2$/Si substrates with pre-patterned Cr/Au electrodes, we identified suitable samples with good geometry using optical microscopy. Note that, we identified the thickness of the flakes through optical contrast, which is a commonly used method for air sensitive samples, see, e.g., ref. [27]. As widely known in the field of two-dimensional materials, samples with various thicknesses show different optical contrast[37]. Therefore, we characterized the thickness of the samples using atomic force microscopy and established the corresponding relationship between optical contrast and thickness beforehand. To preserve the intrinsic properties of the compound and minimize environmental effects, we encapsulated the samples using thin polymethyl methacrylate films with thicknesses around ~50 nm, which ensured that the samples on the devices were never exposed to air directly. All sample fabrication processes were performed in a glovebox with a gas purification system (<1 ppm of O$_2$ and H$_2$O).

We chose to employ small, uniformly thick flakes of four and six atomic layers of $\alpha$-Bi$_4$Br$_4$ in order to systematically investigate the Aharonov-Bohm oscillations in samples with specific dimensions, allowing for a precise study of the quantum coherent transport behavior of hinge modes. These flakes possess a regular shape, ensuring the straightforward extraction of the magnetic flux area for a given magnetic field direction, and they were chosen to be of a small size to minimize the possibility of exceeding the phase coherent diffusion length. Nevertheless, we do not anticipate any hindrance to observing Aharonov-Bohm oscillations in thicker flakes, provided that the perimeter of the sample's *bc* plane remains within the phase coherent diffusion length regime. We believe that future investigations focusing on the thickness dependence of the Aharonov-Bohm effect in $\alpha$-Bi$_4$Br$_4$ devices will provide valuable insights.

### IV. Sample characterizations

We captured the optical images using an Olympus BX 53M microscope. To determine the thickness of the flakes, we employed atomic force microscopy. Atomic force microscopy images were taken with Bruker Dimension Icon3 in tapping mode.

To gain insights into the crystallographic orientations of the exfoliated flakes, we conducted structural analysis using electron diffraction measurements. Notably, the flakes obtained through mechanical exfoliation on the (001) surface consistently exhibit a rectangular shape with a longer dimension along a specific direction, reminiscent of the shape of the bulk single crystal; this is evident in all of the optical microscopy images of the devices featured in the main text, as well as in the scanning electron microscopy image of a relatively thicker flake (~100 nm thick) shown in Extended Fig. 2**a**. Note that the electron diffraction experiments were performed on this relatively thicker flake. Conducting such measurements on very thin flakes poses technical challenges, as they tend to fully oxidize during sample preparation for electron diffraction. In Extended Fig. 2**b**, we display the prepared cross-sectional transmission electron microscopy thin lamella, which was obtained through focused ion beam cutting that was required for performing the electron diffraction experiments. The inset of Extended Fig. 2**b** provides the electron diffraction patterns acquired at $T = 290$ K. Our diffraction analysis confirms that the long and short axes of the thin flake align along the *b*-axis [010] and *a*-axis [100], respectively, while the surface corresponds to the *ab* plane (001). These observations are consistent with the expectations based on the crystal structure of $\alpha$-Bi$_4$Br$_4$[16,18,19,25,26], where the long axis is anticipated to align along the atomic chain direction. Consequently, in our mechanically exfoliated flakes showcased in the main text, we have designated the direction of the long edge as the crystallographic *b*-axis, and the orthogonal direction on the (001) surface as the *a*-axis. This methodology for determining crystallographic orientations has been effective for our flakes, which are markedly rectangular.

Furthermore, to ensure that the Raman shift from the exfoliated flake matched that of the bulk $\alpha$-Bi$_4$Br$_4$ crystal, we performed Raman spectroscopy on representative flakes of the same optical contrast (See Extended Fig. 3). Raman spectra were collected using Horiba-Jobin-Yvon Raman system under 532 nm laser excitation with a power of 0.1 mW. The Si peak at 520.7 cm$^{-1}$ was used as the reference for the calibration in the data analysis.

### V. Electrical transport measurements

Transport measurements were conducted using an Oxford Heliox system with a temperature range of 0.3 K to 8 T. For measurements under higher magnetic fields up to 18 T, a dilution refrigerator with a base temperature of 30 mK was utilized at the National High Magnetic Field Laboratory in Tallahassee, Florida, USA. To ensure precise sample alignment with the external magnetic field, the devices were mounted on a rotator that allows in situ sample rotation. Multiple devices were prepared and measured over multiple measurement runs to ensure reproducibility of the results.

### VI. Scanning tunneling microscopy measurements

Scanning tunneling microscopy measurements were performed on freshly cleaved samples. To obtain atomically resolved, pristine surfaces, single crystals were mechanically cleaved in situ at 77 K under ultra-high vacuum conditions. The freshly cleaved samples were immediately inserted into the microscope head, already at the $^4$He base temperature (4.2 K). For each cleaved crystal, surface areas over $5 \times 5$ μm$^2$ were explored to find atomically flat surfaces. Topographic images were acquired using a Unisoku Ir/Pt tip in the constant current mode. Tunnelling conductance spectra were obtained using standard lock-in amplifier techniques with a lock-in frequency of 977 Hz. The tunneling junction set up and the modulation voltage for lock-in detection are mentioned in the corresponding figure captions.

### VII. Scanning tunneling microscopy evidence of hinge modes in four-atomic-layer-thick step edges

In this section, to reinforce the argument that the observed Aharonov-Bohm effect in the transport experiment is indeed a result of the transport of hinge state carriers (as illustrated in Fig. 1**c**), we present additional evidence supporting the existence of these states in four-atomic-layer-thick step edges of $\alpha$-Bi$_4$Br$_4$ crystals. To achieve this, we employed scanning tunneling microscopy, which provides a direct, atomic-scale visualization of localized states along atomic step edges with high spatial and energy resolution[38]. This technique has been effectively used to identify topological edge states in many quantum materials[19,38-50], including $\alpha$-Bi$_4$Br$_4$[19]. After conducting a thorough scan of a freshly cleaved $\alpha$-Bi$_4$Br$_4$ crystal, we find two four-layer-thick atomic step edges along the *b* axis with opposite orientations identified by the topographic images in Extended Fig. 4**a**. The two sides of the step edge exhibit the A-type surface, which is revealed by magnified topographic images in Extended Fig. 4**b**. The thickness of the step edges is equal to four atomic layers, as indicated by the height profiles around the two step edges displayed in Extended Fig. 4**c**. Examining the tunneling spectrum (d$I$/d$V$) taken away from the step edges (purple curves in Extended Fig. 4**d**), we observe a large insulating gap of $\simeq 260$ meV, defined as the energy difference between the conduction and valence band edges[19]. We note that the energy gap value may be subject to band bending induced by the probe tip[51,52], which can lead to an overestimation of the spectral gap size acquired from tunneling spectroscopy in insulators or semiconductors[53]. It is important to note that in $\alpha$-Bi$_4$Br$_4$, the Fermi level is located within the energy gap[19]. Nevertheless, it appears that the conduction band edge is near the Fermi level which could lead to the generation of thermally activated carriers at temperatures above $T \simeq 20$ K, as suggested by a resistance drop in the temperature dependent resistance measurements displayed in Extended Fig. 8 (see Section XII in Methods).

Most notably, d$I$/d$V$ taken at the two four-layer step edges, as shown in Extended Fig. 4**d**, reveal significantly different behaviors. The left step edge features a gapless, in-gap state (orange curves), whereas the right step edge spectrum (green curves) shows an energy gap (Fermi energy lies within the gap). This inversion asymmetric behavior is further demonstrated in the spatially resolved d$I$/d$V$ map taken at the Fermi energy (Extended Fig. 4**e**). It is evident from Extended Fig. 4**e** that the edge state mainly appears at the left step edge, while for the right step edge d$I$/d$V$ is substantially suppressed compared to that on the left step edge. This asymmetric behavior is reproducible across various samples and tips; see Extended Fig. 5 for a second set of topographic and spectroscopic data acquired using a different sample and tip. It can be elegantly explained through the quantum hybridization of the edge modes. Notably, monolayer $\alpha$-Bi$_4$Br$_4$ is a quantum spin Hall insulator, featuring Z$_2$ topological edge modes[16,19]. The hybridization of two such Z$_2$ edge modes is destructive and opens an energy gap. Because the inversion center in $\alpha$-Bi$_4$Br$_4$ is in the monolayer and even-numbered layers lack inversion symmetry[16], the left and right four-layer edges have different geometries (Extended Fig. 4**f**). There are two types of AB edges: those with facing angles larger than 180°, which lead to weaker hybridization, and those with facing angles smaller than 180°, resulting in stronger hybridization[16,19]. This interlayer edge state hybridization has been identified as a key building block of bulk higher-order topological insulators[12-15] that exhibit helical hinge states[16]. In our case, as illustrated in Fig. 4**f**, the asymmetric hybridization ensures that the left four-layer edge carries hinge states on top and bottom hinges, while the right four-layer edge does not carry any hinge state. Consequently, we obtain a hinge state configuration consistent with the one presented in Fig. 1**c**.

In this context, it is worth noting that as of now, there is no scanning tunneling microscopy evidence available for the hinge mode along the $c$-axis. The absence of such experimental data is primarily due to the challenges associated with cleaving the crystal along either (100) or (010) planes for tunneling measurements. Since the preferred cleavage plane for $\alpha$-Bi$_4$Br$_4$ is along (001), as confirmed by X-ray diffraction measurements (Section II), cleaving the sample in other planes poses significant challenges. However, as expounded upon in Sections VIII and IX through symmetry arguments, the formation of a hinge mode along the $c$-axis is crucial for connecting the hinge modes along the $b$-axis direction (as observed in scanning tunneling microscopy experiments) on the top and bottom hinges, enabling the transfer of a spectral flow[16]. Indeed, in a finite-sized sample, such a hinge state is expected to emerge from the gapped Dirac surface state at the (010) surface. Furthermore, as also elaborated upon in Section IX, the Aharonov-Bohm effect arising from the hinge mode carriers circumnavigating along the $bc$ plane implies the existence of gapless hinge modes along the $c$-axis direction as well.

Lastly, it is worth mentioning that while scanning tunneling microscopy visualizes the hinge modes on a scale of tens of nanometers, these modes may extend over much larger lengths, even in the presence of defects. In fact, it is expected that there will be defects on the hinges of the samples used for the transport measurements. However, due to the topological nature of the hinge modes, they are resilient against one-dimensional Anderson localization for arbitrarily strong nonmagnetic disorder. While the disorder may re-route the modes, it will not break their spectral flow and their inherent one-dimensional conductivity. It is important to note that this local protection of the hinge modes persists even if the disorder breaks the inversion symmetry used to characterize the higher order topology in the bulk (which is, in fact, likely to be the case for the geometry of the sample subjected to disorder).

### VIII. Higher-order band topology of $Bi_4Br_4$

Three-dimensional $\alpha$-$Bi_4Br_4$ displays two distinct types of topological boundary states- higher-order hinge states and surface Dirac cones. The first type, namely the hinge states due to the higher-order band topology, is protected by time-reversal and inversion symmetries. $\alpha$-$Bi_4Br_4$'s symmetry indicators, $Z_{2,2,2,4} = \{0,0,0,2\}$[54], indicate that it is neither a strong topological insulator nor a weak topological insulator. Instead, it features a double band inversion in the bulk and a single protected helical hinge state on an inversion-symmetric path on the surface of any inversion-symmetric crystal. The specific hinges featuring this state depend on the surface termination and cannot be predicted from the bulk electronic structure and its topological invariants[16]. Moreover, the helical hinge state can be protected by time-reversal symmetry alone, indicating its ability to survive under weak inversion symmetry breaking and in crystals with inversion-asymmetric terminations[16]. The second type of topological boundary state that the three-dimensional $\alpha$-$Bi_4Br_4$ exhibits is surface Dirac cones. It has a nontrivial $Z_2$ topological invariant protected by the $C_2$ rotation symmetry around the [010] axis, resulting in a single band inversion in each of the two $C_2$ invariant subspaces[54]. This guarantees the existence of two gapless Dirac cones on any surface that preserves the $C_2$ rotation symmetry, i.e., the (010) and (0-10) surfaces. When the $C_2$ rotation symmetry is broken at these surfaces, the surface states become gapped[16]. Therefore, an inversion- and $C_2$- symmetric rod-like crystal with two capping surfaces normal to the [010] direction would feature both, helical hinge states along the [010] direction and gapless Dirac cones at the (010) and (0-10) surfaces. The two topological invariants are not completely independent, as a capping surface that is not gapless would not allow for the hinge states to connect in a way that preserves the $C_2$ symmetry, given that they need to form a singly connected loop due to the spectral flow they carry[16]. Notably, when the $C_2$ rotation symmetry is broken at the capping surfaces, the surface states become gapped, and their 'residue' forms the gapless hinge states connecting the original hinge states[16]. Importantly, the termination of $\alpha$-$Bi_4Br_4$ surfaces has a significant effect on its hinge-state profile[16]. In a sample with uniform cuts along the (001), (00-1), (100), and (-100) surfaces, there is always one hinge at the top surface and one at the bottom surface that host a helical hinge state. Odd-layer systems are inversion-symmetric, and these two special hinges are located on the opposite side surfaces[16]. However, in even-layer systems such as the four-layer flakes studied in this work, the inversion symmetry is broken, and the two special hinges are located on the same side surface[16,19], either the (100) or (-100) surface, as illustrated in Extended Fig. 6**a**. The even-layer $\alpha$-$Bi_4Br_4$ also features gapless Dirac cones at the (010) and (0-10) surfaces protected by the $C_2$ rotation symmetry around the [010] axis. As discussed earlier, if the $C_2$ rotation symmetry is broken at the capping surfaces, the surface states become gapped, and their residual states form the gapless hinge states connecting the original hinge states, as illustrated in Extended Fig. 6**b** (also in Fig. 1**c**). We refer to Section IX for further discussion regarding the formation of hinge states along the edges of one of the (100) surfaces of a finite-sized $\alpha$-$Bi_4Br_4$ flake used in our transport experiments.

### IX. Theoretical calculations of four-layer $\alpha$-$Bi_4Br_4$

To analyze the electronic properties of $\alpha$-$Bi_4Br_4$, and determine its topological character, we performed first-principles calculations using the VASP code[55, 56]. To obtain more precise information on band gaps and inversions, we utilized the Heyd-Scuseria-Ernzerhof hybrid functional method[57]. We also employed the Wannier90 code[58] to construct maximally localized Wannier functions for the *p*-orbitals of both Bi and Br. These Wannier functions were then utilized to construct ab initio tight-binding model[16] for the material. This enabled us to analyze the electronic band structure and better understand the topological boundary states of $\alpha$-Bi$_4$Br$_4$. As discussed in Section VIII, $\alpha$-Bi$_4$Br$_4$, as a higher-order topological insulator, exhibits a hinge-state profile that is dependent on the termination of its surfaces[16]. The inversion asymmetry leads to the two helical hinge states being located at the two hinges of the same side surface, either the (100) or (-100) surface, as demonstrated by our calculations presented in Extended Figs. 7**a, b**. The gapped (100) and (-100) side surface states are represented by orange bands, while the two purple bands indicate the localization of the two helical hinge states at the two hinges of the (100) side surface. These calculations confirm the presence of well-defined hinge states within the bulk and surface gaps in four-layer $\alpha$-Bi$_4$Br$_4$.

Upon comparing our theoretical calculations with tunneling spectroscopy measurements (see Extended Figs. 4 and 5), we observe a subtle difference: the tunneling spectroscopy data shows a gapless state on the left edge, whereas our theoretical calculations for a four-layer system (as depicted in Extended Fig. 7) indicate hybridization of the left bottom and left top states, leading to a small gap of ~ 4 meV. Such hybridization is expected in a higher-order topological insulator. In this context, it is important to note that the complete gaplessness of the hinge modes is theoretically anticipated only when the hinges are separated by a thermodynamically large distance. Therefore, in a nanoribbon structure with only four atomic layers (which falls short of meeting this criteria), it is theoretically anticipated that the hinge states would exhibit a finite gap. The size of this gap decreases exponentially as the number of layers increases[59]. Consequently, the gapless nature of the hinge modes in a four-atomic layer scenario is not guaranteed by the bulk topology. However, depending on material parameters, such as the surface correlation length, the hinge mode may remain gapless down to very few layer heights. For example, as demonstrated in bismuth[40,60], a pure higher-order topological material, the edge state has been observed on steps with the height of a bismuth bilayer, without exhibiting a gap[40,60]. Nevertheless, it is important to emphasize that, in transport experiments, this gap (if present) will generally impact the transport properties only if it develops precisely at, or very close to, the Fermi energy. Therefore, in our transport sample, even if a gap were to develop, if it occurs at an energy level distant from the Fermi energy, it is unlikely to affect the observed Aharonov-Bohm effect.

Finally, it is worth noting that our theoretical calculations primarily emphasize the hinge modes along the *b*-axis and do not explicitly address the hinge modes forming along the *c*-axis to create a singly connected loop on the (100) surface, a configuration necessary for carrying spectral flow (as depicted in Fig. 1**c**). To elucidate the formation of such a hinge mode loop, we present the following symmetry-based arguments. As elaborated in Section VIII, in an ideal three-dimensional $\alpha$-Bi$_4$Br$_4$ crystal possessing a nontrivial $Z_2$ topological invariant protected by the $C_2$ rotation symmetry around the [010] axis, there should be two topologically gapless Dirac cones on any surface preserving this $C_2$ rotation symmetry, namely the 'perfect' (010) and (0-10) surfaces[16]. However, it is crucial to acknowledge that when the $C_2$ rotation symmetry is broken at these surfaces, the surface states become gapped. Most likely, this occurs in our finite-sized $\alpha$-Bi$_4$Br$_4$ flakes used for transport measurements due to their finite length along the *b*-axis. Furthermore, as a crystal termination perpendicular to the atomic chains along the *b*-axis in a finite-sized sample, the (010) surface is not a natural cleavage plane, and the $C_2$ rotation symmetry is not expected to be preserved. Therefore, strictly speaking, the (010) surface is always expected to exhibit a gap. Though, in cases where the $C_2$ rotation symmetry is 'on average' preserved, the gap on the (010) surface may be very small in the limit of large (010) surface area. Nevertheless, in the four-layer (or six-layer) system, (1) the $C_2$ rotation symmetry is clearly broken (due to the even number of layers)[16], and (2) the number of stacked layers is small, implying that the (010) surface gap is expected to be clearly opened. Now, if a capping surface is not gapless, there is no way to connect the top and bottom hinge states along the *b*-axis while preserving the $C_2$ symmetry, a requirement since the hinge states must form a singly connected loop to carry the spectral flow[16].

In our finite-sized samples, both the (001) and (100) surfaces lack gapless surface states. Additionally, the Dirac surface state at the (010) surface gaps out due to the breaking of $C_2$ rotation symmetry (as discussed above). Hence, the most plausible scenario for forming a singly connected hinge mode loop is that the "residue" of the gapped Dirac surface state on the (010) surface brings about conducting channels localized on the *c*-axis hinges. Consequently, the gapless hinge states along the *c*-axis are expected to form and connect the original hinge states (as discussed in ref. [16]). This results in a singly connected loop on the (100) surface that effectively carries the spectral flow, as illustrated in Extended Fig. 6**b** (and Fig. 1**c**).

Given these considerations, capturing the *c*-axis hinge modes in theoretical calculations requires modeling 'imperfect', non-cleavable (010) and (0-10) surfaces. Such a calculation is exceedingly complex, and we anticipate that its outcomes may not yield further insights beyond the symmetry-based arguments presented here, as they sensitively depend on the exact termination chosen for the modeling.

Having provided a rationale for the existence of *c*-axis hinge modes in our α-Bi$_4$Br$_4$ flakes, we emphasize that, experimentally, the manifestation of the Aharonov-Bohm effect requires electrons to traverse the sample in a closed loop to induce interference. Therefore, our experimental observation of the Aharonov-Bohm effect inherently supports the presence of such a closed loop. Furthermore, our angular-dependent Aharonov-Bohm effect measurements provide compelling evidence indicating that this closed loop resides on the *bc* plane, with the extracted area over which the carriers maintain phase coherence, closely resembling the flux area of the *bc* plane. Hence, to account for the hinge modes forming a closed loop on the *bc* plane, the most plausible scenario entails these hinge states forming along the boundaries of the *bc* plane, specifically along the *b*- and *c*-axes of the α-Bi$_4$Br$_4$ flakes.

### X.   Magnetic field effect on the hinge state

In this section, we explore the influence of a magnetic field on the hinge state. The impact of the magnetic field can be elucidated as follows: The intrinsic gapless nature of the helical edge state, i.e., the Kramers degeneracy at the Dirac point, is protected by the time-reversal symmetry. When this symmetry is broken via the application of a magnetic field, a Zeeman energy gap may emerge at the Dirac point. Let us consider a simplified model Hamiltonian for the helical edge state: $H_0 = v\hbar k_y \sigma_y$, where $\sigma_i$ ($i = x, y, z$) represent the Pauli matrices, $v$ is the Dirac velocity, and $\hbar$ denotes the Planck constant. The corresponding dispersions are $E_\pm = \pm v\hbar k_y$, and the two bands linearly cross at the Dirac point at $k_y = 0$. When the magnetic field is applied along the *z*-axis, the resulting Zeeman term can be modeled as $H_\perp = \alpha_z B_z \sigma_z$, where $\alpha_i$ ($i = x, y, z$) are coefficients related to the Landé g-factor. In this case, the dispersions become $E_\pm = \pm(v^2\hbar^2 k_y^2 + \alpha_z^2 B_z^2)^{1/2}$, i.e., the Dirac point (at $k_y = 0$) becomes gapped. We can provide a rough estimate for the gap size based on field-dependent spectra collected at a monolayer step edge, as reported in ref. [19], which shows a Zeeman gap opening rate on the order of a meV/T. In our transport experiments, we observed the Aharonov-Bohm effect up to a magnetic field of 18 T. Assuming that the Zeeman gap opening rate observed for a monolayer step edge in ref. [19] also applies to the four- (or six-) layer thick α-Bi$_4$Br$_4$ flakes used in our experiments, a field of 18 T would induce a $\Delta \sim 50$ meV Zeeman gap (over a 260 meV bulk band gap) at the Dirac point at $k_y = 0$. While the exact energy location of the original Dirac points in the α-Bi$_4$Br$_4$ flakes employed for our transport measurements remains unknown, and may not necessarily align with the theoretically calculated counterpart, it is conceivable that the Dirac point is situated beyond $\Delta/2$, i.e., 25 meV away from the Fermi energy (over the energy scale of 260 meV), and it may not even lie within the bulk or surface gap. Given that the transport properties are primarily governed by carriers on the Fermi energy, in such a scenario, it is plausible that this magnetic field-induced gap would exert minimal influence on the transport properties or the Aharonov-Bohm effect up to a magnetic field of 18 T.

### XI. An overview of the Aharonov-Bohm effect in mesoscopic electron transport:

In this section, we contextualize our findings by first reviewing relevant prior works, and then discuss our observations based on this background.

Experimental investigations into the Aharonov–Bohm effect in mesoscopic electron transport gained prominence in the 1980s. Sharvin and Sharvin conducted pioneering experiments involving a slender magnesium cylinder deposited around a micron-thin quartz filament[61]. When a magnetic field was applied along the cylinder's axis, they observed periodic oscillations in the magnetoconductance with only $h/2e$ (normalized by the cross-sectional area of the cylinder) periodicity, whereas the $h/e$ periodicity was absent. A similar oscillation pattern was later observed in carbon nanotubes[33]. This phenomenon may be attributed to the persistent destructive interference between time-reversed pairs of paths, which inhibits the fundamental $h/e$ period even at higher magnetic fields. This effect, which is akin to weak antilocalization, was analyzed theoretically by Altshuler, Aronov, and Spivak: If many such pairs contribute to transport with uncorrelated zero-field phases, the $h/e$ oscillations average out[32,62]. However, the $h/2e$ periodicity contains a significant contribution of time-reversed paths that have the same relative phase, making them more robust against averaging. Subsequent studies on various materials, including metals [34, 35,63,64], semiconductors [65-67], and semimetals (e.g., graphene[36]), conducted using ring-shaped planar geometries (where the diameter of the ring does not exceed the phase coherent diffusion length, ensuring that the entire sample retains quantum mechanical coherence), observed the $h/e$ periodicity.

The field of topological materials witnessed its first observation of the Aharonov–Bohm effect in 2009 when Peng et al.[4] observed it in a strong topological insulator nanoribbon. These nanoribbons can be envisaged as hollow metallic cylinders with surface transport channels. Consequently, the quantum interference of surface carriers following various paths within this hollow structure can lead to oscillatory conductance patterns in response to magnetic flux[62, 68]. Peng et al.[4] observed fundamental $h/e$ periods along with a much weaker $h/2e$ period. It was argued that the lower degeneracy of the topological surface states removes the self-averaging effect, thus bringing about the fundamental $h/e$ period[62]. Furthermore, the absence of $h/2e$ periodicity at low fields in this topological insulator nanoribbon was attributed to the absence of weak anti-localization behavior for integer multiples of the $h/2e$ flux[4]. Note that the $h/2e$ oscillation was present at higher magnetic fields[4]. Subsequent studies on the Aharonov–Bohm oscillations in three-dimensional topological insulators and topological semimetals, involving nanoribbons, nanowires, or ring-shaped geometries, reported similar oscillation periodicities, indicating surface transport[5-10].

Interestingly, in another topological material, 1T′−MoTe$_2$, only $h/2e$ periodicity is observed[17]. 1T′−MoTe$_2$ has the same higher-order topological band index as $\alpha$-Bi$_4$Br$_4$, but it is metallic in the bulk and on the surface, and hence not a bulk higher-order topological *insulator*. The observed periodicity may be due to this (predominantly surface) metallicity and the associated strong chemical potential fluctuations induced by disorder[68,69]. These states are non-topological, and therefore, one expects them to exhibit the same self-averaging behavior as non-topological cylinders. In contrast, in our study on $\alpha$-Bi$_4$Br$_4$, we observe both $h/e$ periodic Aharonov–Bohm and $h/2e$ periodic

Altshuler–Aronov–Spivak oscillations, with the primary *h/e* periodicity dominating over the *h/2e* periodicity. This observation indicates the absence, or at least the suppression, of self-averaging, as would be expected from the topological nature of the associated hinge modes, as well as the presence of weak anti-localization.

### XII. Temperature dependent resistance measurements

In this section, we present the temperature-dependent resistance data obtained from Sample 1. This data, shown in Extended Fig. 8, suggests the presence of hinge states in $\alpha$-Bi$_4$Br$_4$. Specifically, we observe that for temperatures exceeding $T \cong 20$ K, an Arrhenius plot of the normalized resistivity, $\frac{R_{xx}}{R_{xx}(T=300\ K)}$, reveals a clear gap, as indicated by the linear behavior shown in the inset of Extended Fig. 8. This activated behavior is suppressed once $T$ is lowered below 20 K leading in the Arrhenius plot to a nearly constant dependence for $\frac{R_{xx}}{R_{xx}(T=300\ K)}$ as a function $1/T$. This observation indicates that the gapless hinge states effectively short circuit the bulk gap by providing additional channels for carrier conduction.

**Note Added:** We recently learned about two related theoretical works[70,71] on interferometry in higher-order topological insulators. They discuss the Aharonov–Bohm interference of topological hinge states.

**Data and materials availability:** All data needed to evaluate the conclusions in the paper are present in the paper. Additional data are available from the corresponding authors upon reasonable request.

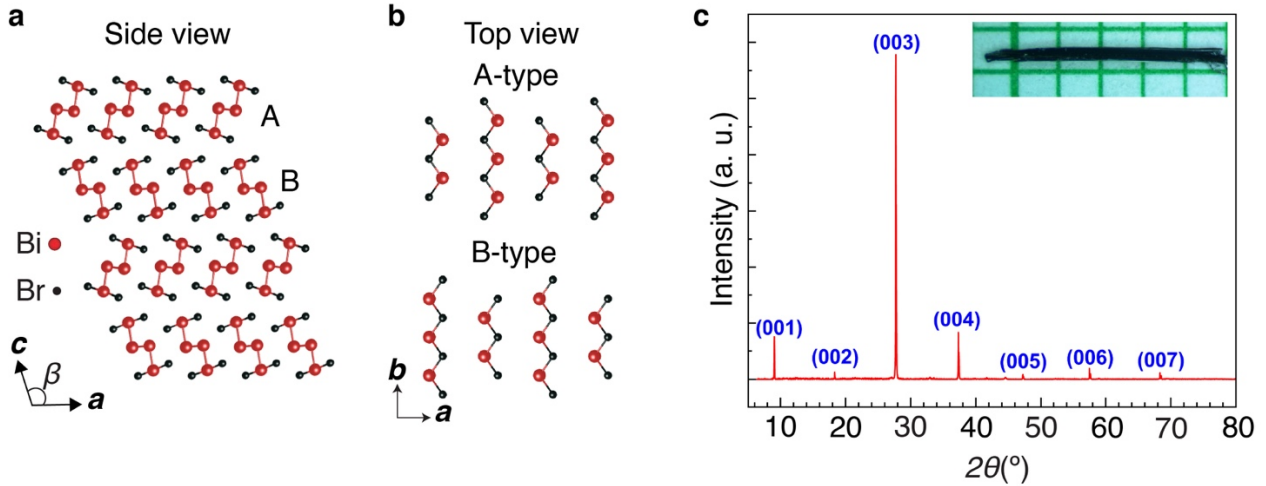

**Extended Fig. 1: Crystal structure and experimental X-ray diffraction data of $\alpha$-Bi$_4$Br$_4$. a,** Side view of the crystallographic structure of $\alpha$-Bi$_4$Br$_4$, showing the interlayer AB stacking. **b**, Top view of the A-type and B-type monolayers. **c**, X-ray diffraction data obtained from a single crystal of $\alpha$-Bi$_4$Br$_4$, featuring the (00l) preferred peaks. The refined lattice parameters are: $a = 13.0667(12)$ Å, $b = 4.3359(4)$ Å, $c = 20.0676(19)$ Å, and $\beta = 107.336(2)°$. The inset displays an optical microscopy image of a typical $\alpha$-Bi$_4$Br$_4$ bulk single crystal, placed on millimeter grid paper, with the top surface corresponding to the (001) plane. This surface was used for the mechanical exfoliation process to fabricate the devices utilized in our transport measurements.

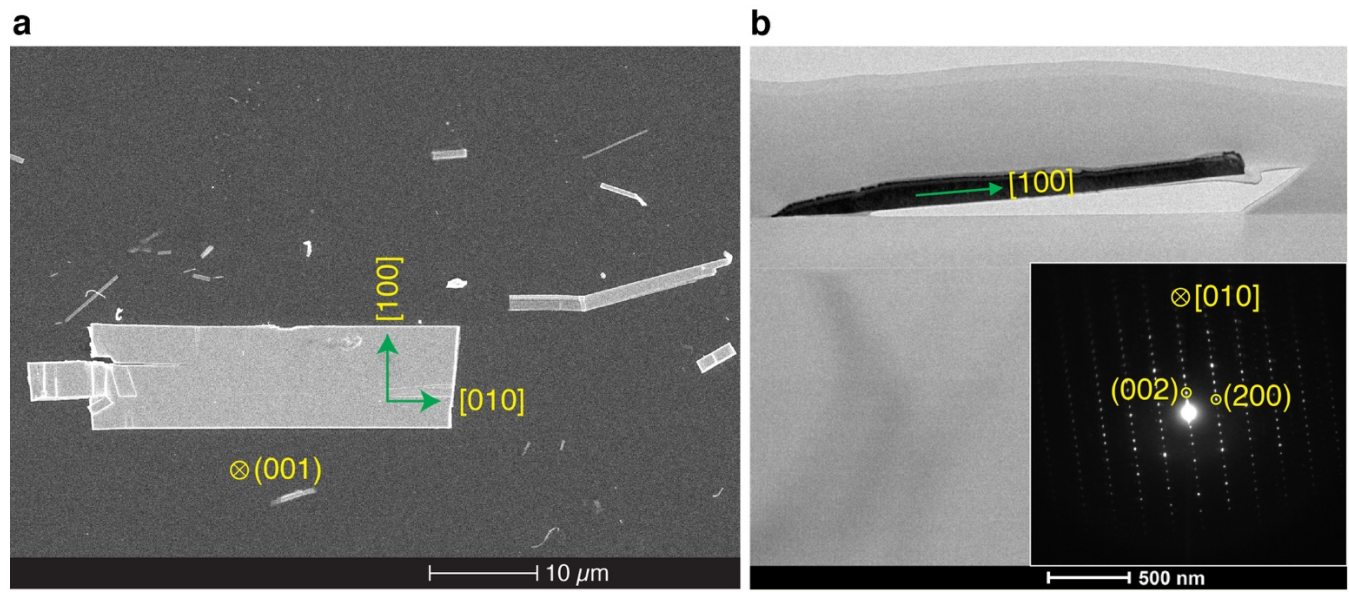

**Extended Fig. 2: Electron diffraction analysis of a mechanically exfoliated α-Bi₄Br₄ flake**. **a**, Scanning electron microscopy image of rectangular-shaped α-Bi₄Br₄ thin flakes on a Si substrate. The crystallographic plane and the axes are indicated for the largest flake used for the diffraction analysis. **b**, Transmission electron microscopy image displaying the cross-section of the thin flake, which is perpendicular to the long axis of the flake. Inset: corresponding electron diffraction pattern from the [010] zone axis.

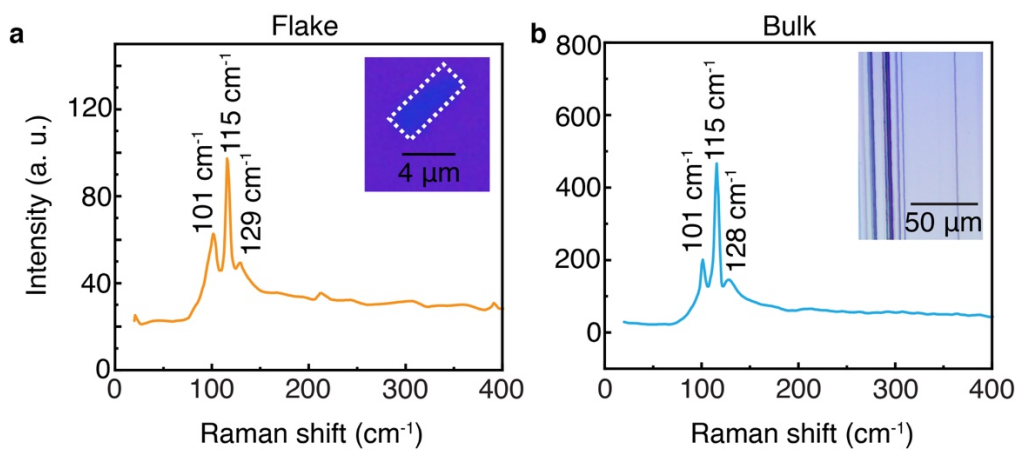

**Extended Fig. 3: Raman spectroscopy of exfoliated and bulk α-Bi₄Br₄. a**, Raman spectrum obtained from a mechanically exfoliated α-Bi₄Br₄ flake consisting of four atomic layers. The inset shows an optical microscopy image of the corresponding α-Bi₄Br₄ flake on a SiO₂/Si substrate, with the flake boundary outlined by a white dashed rectangle. **b**, Raman spectrum of bulk α-Bi₄Br₄, accompanied by the corresponding optical microscopy image in the inset. The Raman shift acquired from the four-layer flake matches the bulk Raman shift, indicating the preservation of the crystalline properties throughout the exfoliation process. The error bar associated with peak identification is approximately 1 cm$^{-1}$.

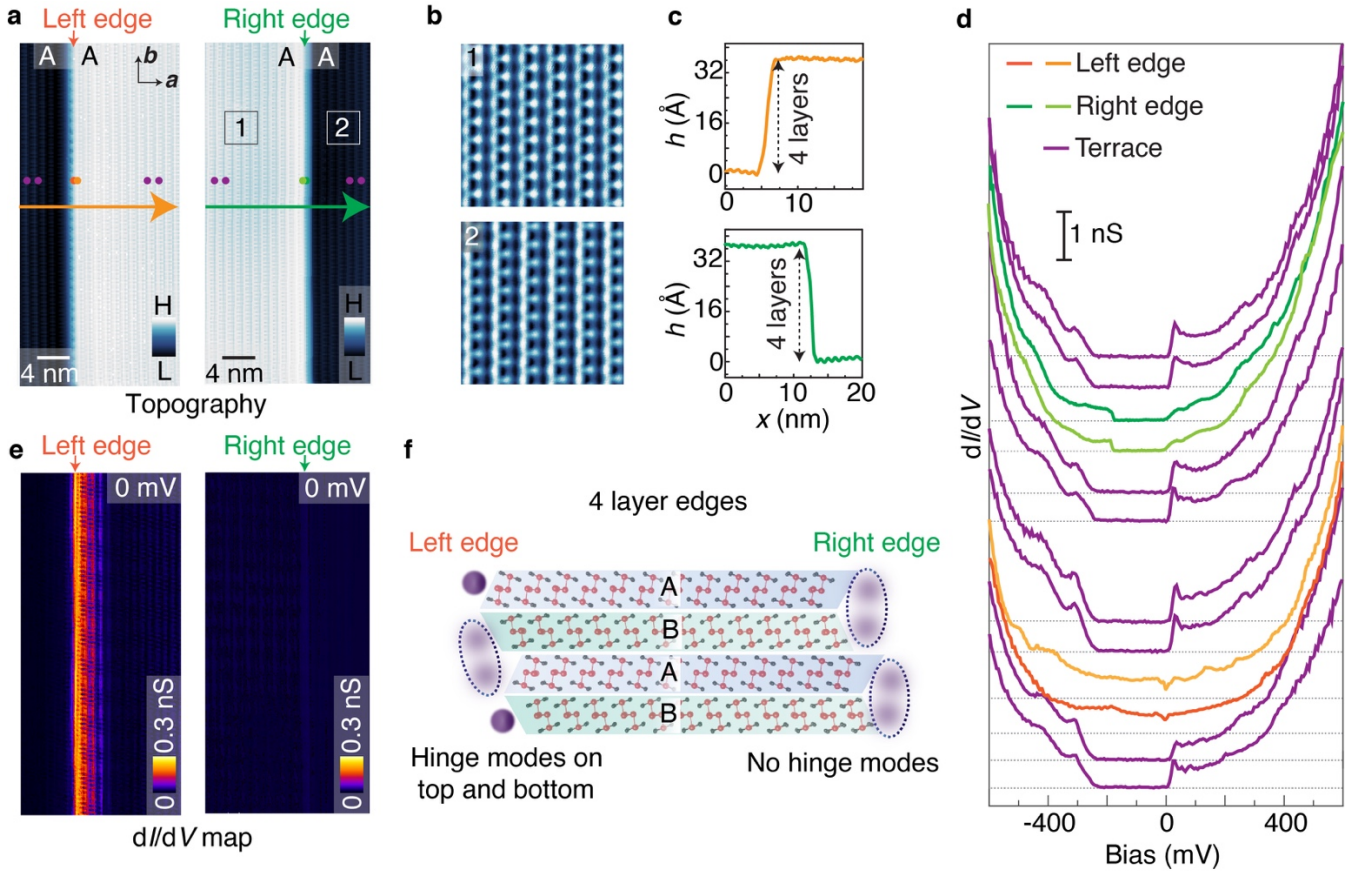

**Extended Fig. 4: Scanning tunneling microscopy evidence for a hinge state in a four-layer step edge. a**, Topographic images of two four-layer atomic step edges of opposite geometric orientations. **b**, Magnified images of the two sides of the four-layer atomic step edges, revealing A type surfaces on both sides. The corresponding regions are marked as 1 and 2 in the large-scale topographic image in panel **a**. **c**, Height profiles of the topographic images in panel **a**, taken perpendicular to the *b*-axis. The corresponding locations are marked on the topographic images in panel **a** with color-coded lines; the scan directions are indicated with arrows. **d**, d$I$/d$V$ spectra shown in color-coding, obtained at the left step edge, right step edge, and regions away from the edge. The positions where these spectra were acquired are marked with matching color-coded dots on the topographic images in panel **a**. The left step edge exhibits a pronounced in-gap state, while the right step edge shows significant suppression of the density of states at the bulk gap. **e**, Differential conductance maps of the two four-layer atomic step edges shown in panel **a**, taken at the Fermi energy ($V = 0$ mV). A pronounced edge state is observed on the left edge whereas on the right edge, no edge state is visible. Tunnelling junction set-up for spectroscopy: $V = -600$ mV, $I = 0.5$ nA, and a root mean square oscillation voltage of 1 mV. Tunnelling junction set-up for d$I$/d$V$ maps: $V = -600$ mV, $I = 0.3$ nA, and a root mean square oscillation voltage of 10 mV. All data were obtained at $T \simeq 4.2$ K. **f**, Schematic of a four-layer edge showing quantum hybridization of the quantum spin Hall edge states for neighboring layers. The edges of A and B layers with a facing angle smaller than 180° exhibit a stronger hybridization of the monolayer edge states (depicted as purple spheres). The hybridization of quantum spin Hall edge state is destructive, as illustrated by the lighter color of the edge states (purple spheres). As a result, the left four-layer configuration hosts hinge states on the top and bottom hinges, while the right configuration lacks hinge states. Therefore, the propagation of the hinge modes across the sample aligns with the schematic representation in Fig. 1c.

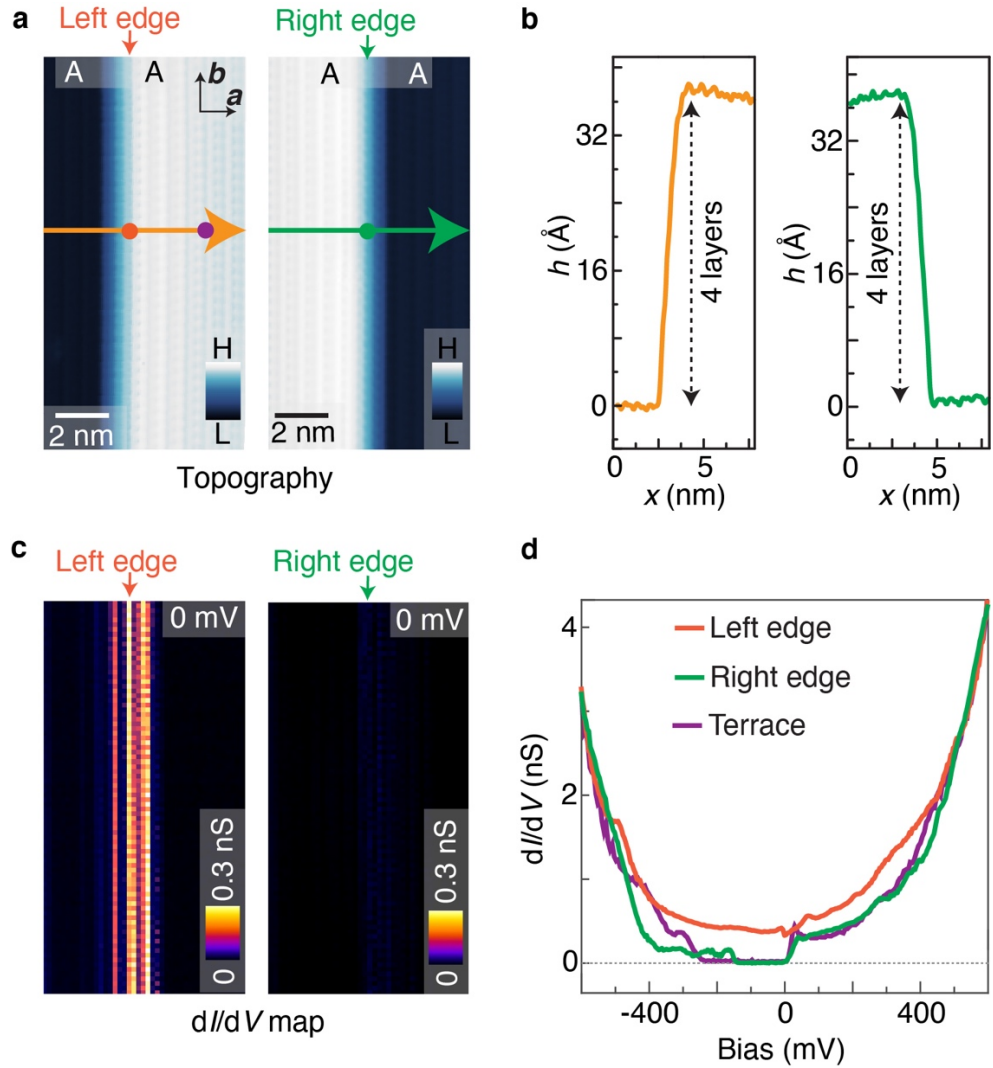

**Extended Fig. 5: Reproducibility of the scanning tunneling microscopy data– evidence for a hinge state in a four-layer step edge, acquired using a different sample and tip. a**, Topographic images of two four-layer atomic step edges with opposite geometric orientations. The terraces on both sides of the step edges exhibit an A-type surface. **b**, Height profiles of the topographic images in panel **a**, taken perpendicularly to the *b*-axis. The corresponding locations are marked on the topographic images in panel **a** with color-coded lines, and the directions of the scans are indicated by arrows. **c**, Differential conductance maps of the two four-layer atomic step edges shown in panel **a**, taken at the Fermi energy ($V = 0$ mV). A pronounced edge state is observed on the left edge, while no edge state is visible on the right edge. **d**, d$I$/d$V$ spectra taken at the left step edge (orange), right step edge (green), and away from the edge (purple), revealing striking differences between the two step edges. Orange and green dots in panel **a** denote the respective positions, on the left and right step edges, where the differential spectra were collected. The left step edge exhibits a pronounced in-gap state, whereas on the right edge, the density of states at the bulk gap is significantly suppressed. Tunnelling junction set-up for d$I$/d$V$ maps: $V = -600$ mV, $I = 0.3$ nA, and a root mean square oscillation voltage of 10 mV. Tunnelling junction set-up for spectroscopy: $V = -600$ mV,

$I = 0.5$ nA, and a root mean square oscillation voltage of 1 mV. All data were obtained at $T \simeq 4.2$ K and are consistent with Extended Fig. 4 results.

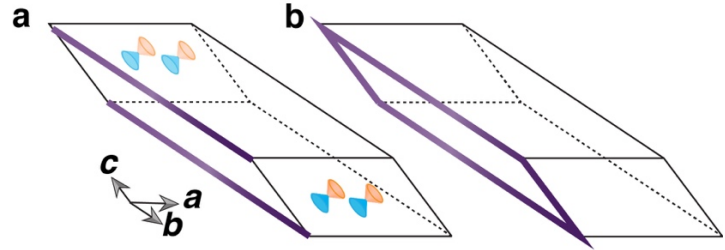

**Extended Fig. 6: Topological boundary states in a $\alpha$-Bi$_4$Br$_4$ nanorod with an even number of layers where the inversion symmetry is broken. a,** Nanorod of $\alpha$-Bi$_4$Br$_4$ featuring helical hinge states resulting from higher-order band topology and two surface Dirac cones on the (010) surface protected by the C$_2$ rotation symmetry around the [010] axis. It is worth noting that the helical hinge states do not require the inversion symmetry and are robust against inversion symmetry breaking[10]. **b,** Propagation paths of the helical hinge states when the C$_2$ symmetry is broken on the (010) surface.

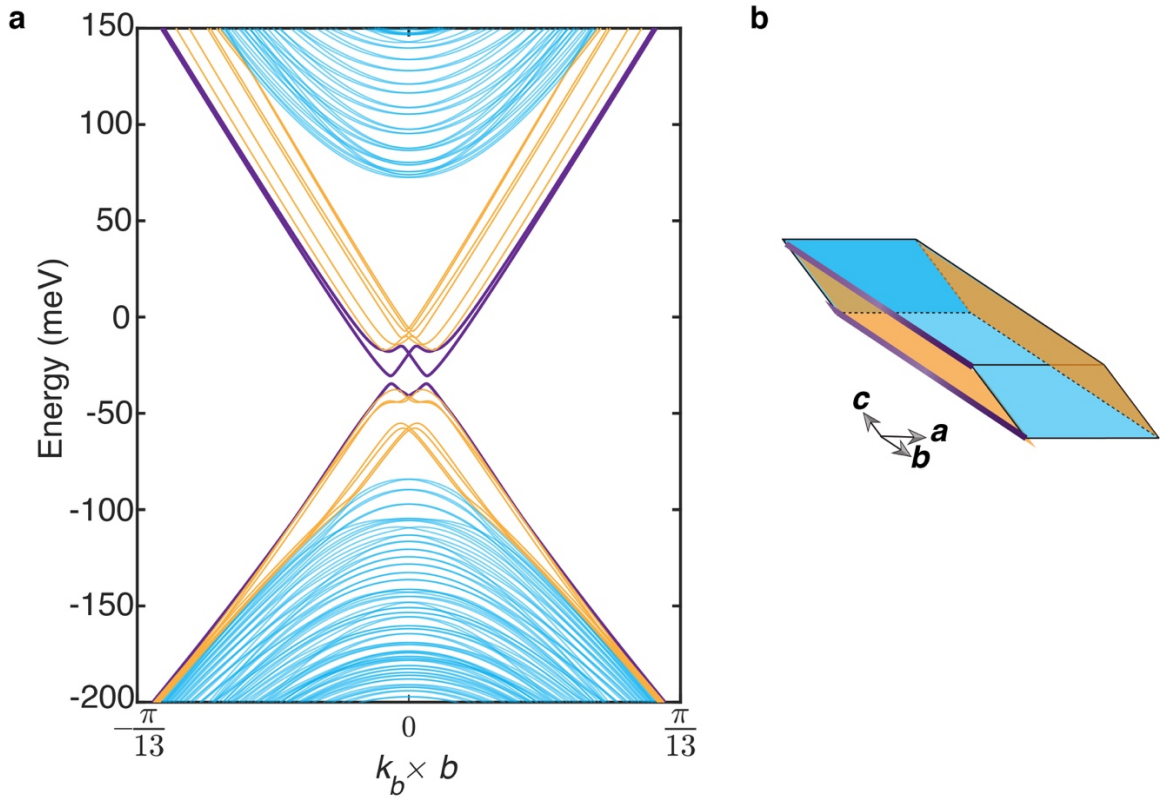

**Extended Fig. 7: Calculated hinge states for a four-layer $\alpha$-Bi$_4$Br$_4$. a,** The edge-projected band structure for a (001) four-layer ribbon on the top surface of $\alpha$-Bi$_4$Br$_4$. Purple bands represent gapless hinge states, cyan bands are from the bulk and (001) surfaces, and orange bands are from the (100) and (-100) side surfaces of the ribbon. Due to the inversion asymmetry inherent in even-layer systems, the bands are singly degenerate at each $k_b$. The ribbon is infinitely long in the *b*-direction and 50-chain wide in the *a*-direction. For a four-layer system, the left bottom and the left top states (depicted in panel **b**) undergo hybridization, resulting in a small gap of 4.1396 meV. Note that the energy gap may be slightly overestimated in this calculation due to its derivation from three-dimensional bulk band structure calculations. In reality, our tunneling spectroscopy measurements (see Fig. 4) reveal the presence of a gapless state on the left edge. Additionally, there exists a quantitative disparity in the positioning of the Fermi energy when comparing the calculation with the tunneling spectroscopy measurements. Nevertheless, the calculation provides a qualitative depiction of well-defined hinge states within the energy gaps of both the bulk and surface states. **b**, Schematic representations in real space illustrating the bulk, surface, and hinge states of a four-layer $\alpha$-Bi$_4$Br$_4$.

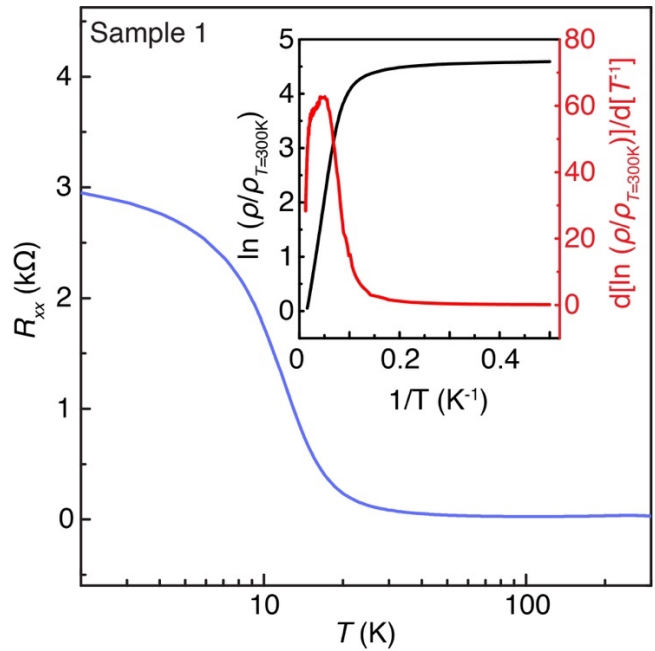

**Extended Fig. 8: Temperature dependence of sample resistance and Arrhenius plot of $\alpha$-Bi$_4$Br$_4$ Sample 1.** Sample resistance, $R_{xx}$ as a function of temperature, $T$ (in Kelvin), exhibiting a sharp rise below $T \simeq 20$ K. Inset: The electrical resistivity $\rho$, normalized to the 300K-value $\rho_{T=300\,K}$, (left axis, black) and its logarithmic derivative (right axis, red) as a function $1/T$.

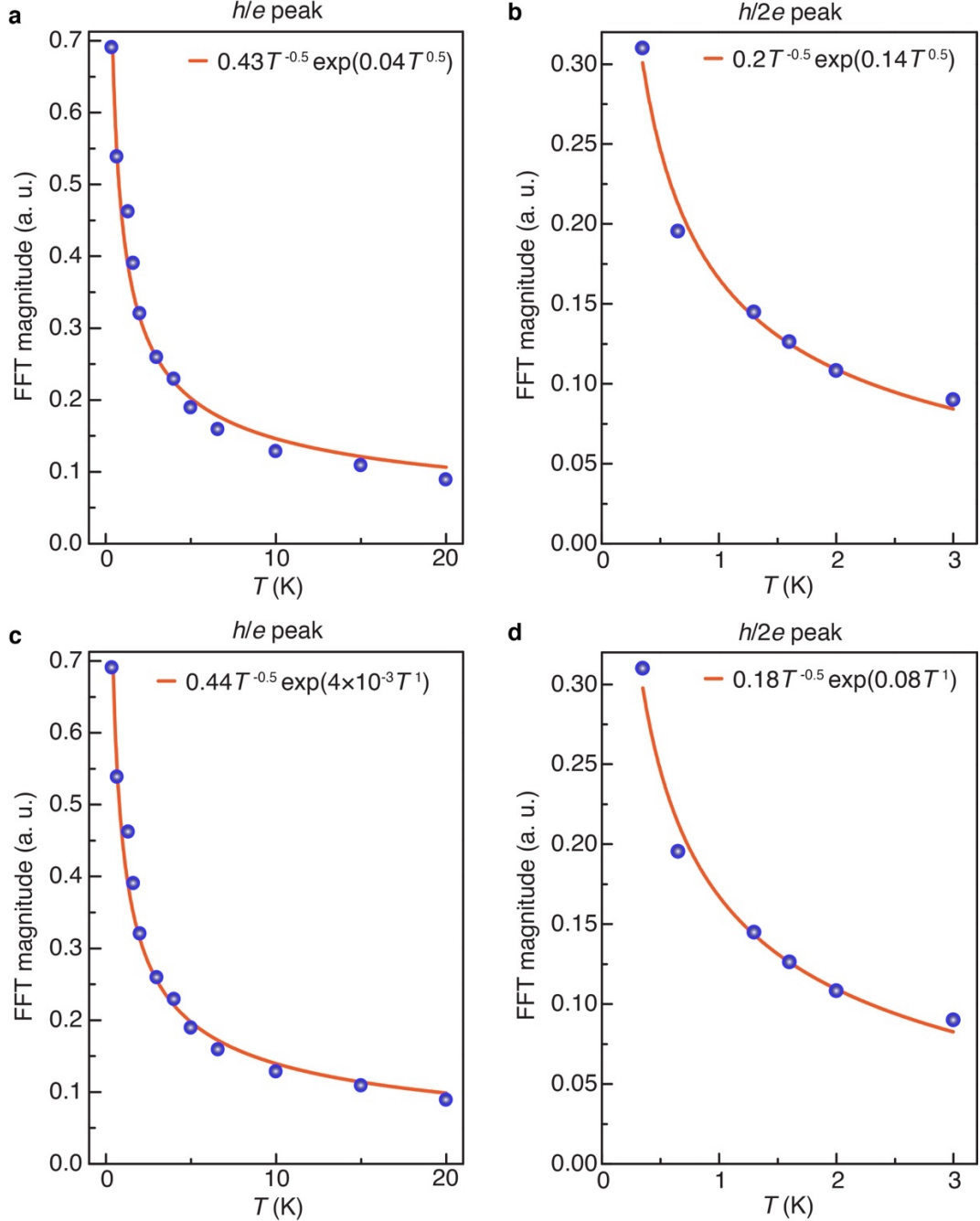

**Extended Fig. 9: Determination of the phase coherent diffusion length from temperature dependent Aharonov-Bohm oscillation data. a**, Temperature dependence of the Fourier transform amplitude of the $h/e$ oscillations. Fitting the data to $AT^{-0.5}\exp(P/L_\phi(T))$ where $L_\phi(T) = BT^{-0.5}$, A and B are the fitting parameters, $P$ is the perimeter of the loop along the $bc$ plane, yields $A = 0.43$ and $\frac{P}{B} = 0.04 \pm 0.02$. The resulting fitting function is represented by the orange curve. **b**, Temperature dependence of the Fourier transform amplitude of the $h/2e$ oscillations. Fitting the data to the same function yields $A = 0.2$ and $\frac{P}{B} = 0.14 \pm 0.06$. The resulting fitting function is represented by the orange curve. **c, d**, Temperature dependence of the Fourier transform amplitude of the $h/e$ and $h/2e$ oscillations, respectively, now fitted considering $L_\phi(T) = BT^{-1}$. The fitting yields $A = 0.44, \frac{P}{B} =$

0.004 ± 0.01 for *h/e* oscillations and $A = 0.18$, $\frac{P}{B} = 0.08 \pm 0.04$ for *h/2e* oscillations. The resulting fitting functions are represented by the orange curves. These fits provide $L_\phi^{AB} = (100 \pm 50)\ T^{-\frac{1}{2}}\ \mu m$ (or $L_\phi^{AB} = (1000 \pm 2500)\ T^{-1} \mu m$) for the Aharonov-Bohm oscillations, and $L_\phi^{AAS} = (60 \pm 26)\ T^{-\frac{1}{2}}\ \mu m$ (or $L_\phi^{AAS} = (100 \pm 50)\ T^{-1}\ \mu m$) for the Altshuler–Aronov–Spivak oscillations.